\documentclass[useAMS,usenatbib]{mn2e}
\usepackage{graphicx} 
\usepackage{longtable}
\usepackage{rotating}
\usepackage{lscape}
\usepackage{color}
\newcommand{\mnras}{MNRAS\,}
\newcommand{\pasp}{PASP\,}

\newcommand{\aap}{A\&A\,}
\newcommand{\aaps}{A\&A\,Supp. \, }
\newcommand{\apj}{ApJ\,}

\newcommand{\aj}{AJ\,}

\newcommand{\apss}{Ap\& SS\,}

\newcommand{\actaa}{Acta Astron.\,}

\title[]{The VMC Survey - X. 
Cepheids, RR Lyrae stars and binaries as  probes of the Magellanic 
System's structure\thanks{Based on 
observations made with VISTA at ESO under program ID 179.B-2003.}}
\author[M. I. Moretti et al.]{M. I. Moretti$^{1,2,3}$\thanks{E-mail:
imoretti@na.astro.it}, G. Clementini$^{2}$, T. Muraveva$^{2}$, V.Ripepi$^{3}$,
J. B. Marquette$^{4}$, 
\newauthor M.-R.L. Cioni$^{5,6,7}$, M. Marconi$^{3}$, L. Girardi$^{8}$,
S. Rubele$^{8}$,  P. Tisserand$^{4,9}$, 
\newauthor R. de Grijs$^{10,11}$,  M. A. T. Groenewegen$^{12}$, R. Guandalini$^{13}$, V. D. Ivanov$^{14}$, 
\newauthor and  J. Th. van Loon$^{15}$
\\
$^{1}$ University of Bologna, Department of Astronomy, via Ranzani 1,
40127, Bologna, Italy \\
$^{2}$ INAF-Osservatorio Astronomico di Bologna, via Ranzani 1,
Bologna, Italy \\ 
$^{3}$ INAF-Osservatorio Astronomico di Capodimonte, via Moiariello
 16, 80131, Naples, Italy \\
$^{4}$ UPMC-CNRS, UMR7095, Institut d'Astrophysique de Paris, 
F-75014, Paris, France\\
$^{5}$ University of Hertfordshire, Physics Astronomy and
 Mathematics, Hatfield AL10 9AB, United Kingdom \\
$^{6}$ Leibnitz-Institut f\"{u}r Astrophysik Potsdam, An der Sternwarte 16, 14482 Potsdam 
Germany \\ 
$^{7}$~Research Fellow of the Alexander von Humboldt Foundation\\
$^{8}$ INAF-Osservatorio Astronomico di Padova, vicolo dell'Osservatorio 5, 35122
Padova, Italy\\
$^{9}$ Research School of Astronomy and Astrophysics, Australian National
University, Cotter Road, Weston Creek ACT 2611, Australia\\
$^{10}$ Kavli Institute for Astronomy and Astrophysics, Peking
University, Yi He Yuan Lu 5, Hai Dian District, Beijing 100871, China\\
$^{11}$Department of Astronomy, Peking University, Yi He Yuan Lu 5, Hai Dian District, Beijing 100871, China\\
$^{12}$Royal Observatory of Belgium, Ringlaan 3, B-1180 Brussels, Belgium\\
$^{13}$Belgium Instituut voor Sterrenkunde, KU Leuven, Celestijnenlaan 200D 2401, 3001, Leuven, Belgium\\
$^{14}$European Southern Observatory, Av. Alonso de Cordoba 3107, Casilla 19, Santiago, Chile\\
$^{15}$Astrophysics Group, Lennard Jones Laboratories, Keele University, ST5 5BG, United Kingdom\\
}
\begin{document}

\date{}

\pagerange{\pageref{firstpage}--\pageref{lastpage}} \pubyear{2002}

\maketitle

\label{firstpage}

\begin{abstract}
The VMC survey is obtaining multi-epoch 
photometry in the $K_\mathrm{s}$  band of the Magellanic System down to a limiting magnitude of 
$K_\mathrm{s} \sim$ 19.3 for individual epoch data. The observations  are 
spaced in time such as to provide optimal sampling of the 
light curves for  RR Lyrae stars and for Cepheids with periods up to 20-30 days. 
We present examples of the $K_\mathrm{s}$-band light curves of Classical Cepheids and RR Lyrae stars we are obtaining from the VMC data and 
outline the strategy we put in place to measure distances and infer the System three-dimensional geometry from the variable stars. 
For this purpose  the near-infrared Period-Luminosity,
Period-Wesenheit, and Period-Luminosity-Colour  relations of the
system RR Lyrae stars and Cepheids are used. We extensively exploit 
the catalogues of the Magellanic Clouds' variable stars provided by the EROS-2 and OGLE~III/IV microlensing surveys. 
By combining these surveys we present  the currently widest-area view of the Large Magellanic Cloud as captured  by the galaxy Cepheids, 
RR Lyrae stars and binaries. This reveals the full extent  of the main structures (bar/s - spiral arms)  that have only been vaguely guessed before.
Our work strengthens the case for a  detailed study of the Large Magellanic Cloud  three-dimensional geometry.
\end{abstract}


\section{Introduction}
The Magellanic Clouds (MCs) are the largest satellites of the Milky
Way (MW) and the nearest external system of interacting galaxies. They 
contain
both old and young stars placing them in a favoured position in the context of
studying the evolution of galaxies.
The MCs are part of a bigger structure, the Magellanic System (MS), formed by the
Large Magellanic Cloud (LMC), the Small Magellanic Cloud (SMC), the Bridge 
 and the Stream. The Bridge and the Stream are mainly formed
by gas, and are the results of the interaction between the two Clouds (e.g. \citealt{Bes12}). 
The 
 LMC is 
  the first step of the extragalactic  distance scale,
hence knowing its  three-dimensional (3-D)  structure holds the key for a proper definition of the entire cosmic distance
scale (see Walker 2012,  and references therein, for a recent review).
The knowledge of the whole MS structure  and of its stellar components is also of crucial importance 
to better understand  the evolution of the two Clouds, the interaction
with the MW and, in turn, to improve our understanding about the formation
of the Galaxy and its satellites. 

Started in November 2009 and expected to  extend beyond the originally
planned $\sim$5 yrs  time span, the 
\emph{VISTA near-infrared $Y, J, K_\mathrm{s}$ 
survey of the Magellanic System}
(VMC\footnote{http://star.herts.ac.uk/$\sim$mcioni/vmc}, P.I.:
M.-R. L. Cioni, see \citealt{Cio11}) 
is studying the star formation history (SFH) and the 3-D structure  of  the MS 
using both constant and variable stars for this purpose. 
The SFH is being recovered by means of the classical CMD-reconstruction method (see \citealt{Rub12}).
 The 3-D geometry  is being inferred from a  number of different  
distance indicators:  the luminosity of the red clump stars, and the Period-Luminosity ($PL$), Period-Luminosity-Colour ($PLC$) and
Period-Wesenheit ($PW$)  relations of the pulsating variable stars
(\citealt{Lea1912}; \citealt{Mad12} and references therein;
\citealt{Lon86}; \citealt{Cop11} and references therein). 
The RR Lyrae stars belong to the oldest stellar component (t $> 10$ Gyrs) in a galaxy  and mainly trace
the galactic halo, whereas the Classical Cepheids (CCs) are the youngest among the radially pulsating 
variables (50-200 Myrs) and mainly reside in star forming regions, galactic bars and spiral arms. These two types of pulsating variables  are thus optimal 
to characterize the spatial structure of MS components with different ages. 
 
The VMC strategy, its main goals and first data are described in \cite{Cio11} (hereafter Paper I),  
first scientific results were 
presented in \cite{Mis11}, \cite{Gul12},  \cite{Rub12}, \cite{Cio13a}, \cite{Tat13}, and \cite{Cio13b}.
First results for the pulsating stars,  based on the VMC  $K_\mathrm{s}$-band light curves,  
were  
presented in \cite{Rip12b}  (hereinafter R12b) for CCs  in the VMC tiles covering the South Ecliptic Pole (SEP),  
and the 30 Doradus (30 Dor)  regions of the LMC, and in  \cite{Rip13}  (hereinafter R13) for  LMC Anomalous Cepheids (ACs).  In two forthcoming 
papers (Moretti et al., in preparation, Muraveva et al., in preparation) we  will present 
results obtained from similar studies of  the LMC RR Lyrae stars (see
\citealt{Rip12a} for preliminary results). 

The microlensing surveys of the MS (\citealt{Wyrzy11a, Wyrzy11b}; \citealt{Tis07}; \citealt{Alc00};
\citealt{Uda97}; see also 
Section~\ref{sec:microlensingSurvey})  increased dramatically the census of
 the MC variable stars by discovering thousands of 
variables in the  two Clouds.  But
an important piece of
information still missing are near-infrared light curves 
for these variables.  The VMC survey is now filling this gap by obtaining 
  $K_\mathrm{s}$ photometry in time-series mode   of the MC variables brighter than $K_\mathrm{s}$ $\sim$ 19.3 mag,  
 with an optimal sampling of the RR Lyrae stars and of  Cepheids with periods shorter than about  20-30 days 
 (saturation limits the observation  of  longer period Cepheids). 
   
In this paper we describe the VMC survey's $K_\mathrm{s}$ time-series data, and outline the procedures we have 
developed, tested and fine-tuned  to derive distances and to study the 
3-D structure of the MS
from the analysis  of the $K_\mathrm{s}$ light curves of  Cepheids and RR Lyrae stars.

The paper is organized as follows. 
Section~\ref{sec:variablesVMC}
describes the type and quality of the VMC data 
 for the variable stars, and compares the VMC sky coverage 
 with that of the 
  microlensing surveys. 
Sections~\ref{sec:ogle30Dor} and ~\ref{sec:EROS_LMC}  
 outline the analysis methods, 
 respectively for the inner and outer LMC  regions  where optical data for
the variable stars are available from different surveys.
We compare in Section~\ref{sec:SFH} our results from the variable
stars with the SFH results while in Section~\ref{sec:LMC_structure} we
discuss the stellar structure of the LMC.
 Finally, Section~\ref{sec:summary}  provides a summary.

\section{Variable stars in the VMC survey}\label{sec:variablesVMC}
\begin{figure}
\begin{centering}
\includegraphics[width=8cm, height=8cm]{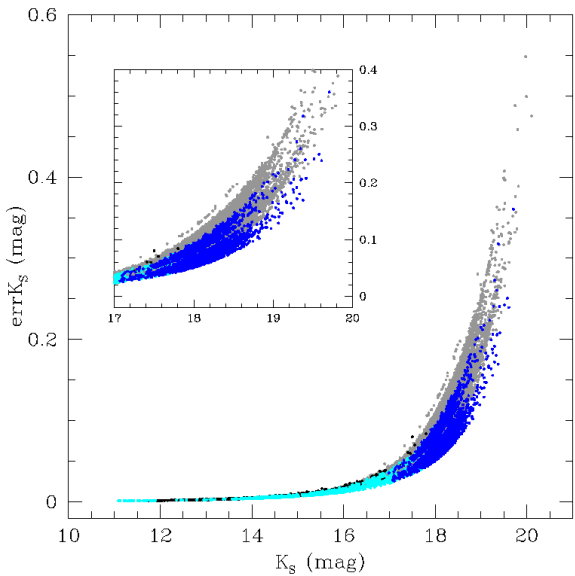}
\caption{Typical errors per single epoch observation from the VSA aperture photometry. 
Grey and blue points correspond to observations of RR Lyrae stars,  black and 
cyan points correspond to observations of CCs   in 
the VMC tiles LMC 6\_6 (30 Dor region) and  8\_3,  respectively.  
The inset shows an enlargement of the RR Lyrae region. See text for
details.}\label{fig:errors} 
\end{centering}
\end{figure}

The VMC survey is imaging $\sim$ 200 $\rm deg^2$ of the MS 
 in $YJK_\mathrm{s}$ ($\lambda_c$=1.02, 1.25, and 2.15 $\mu$m, respectively) reaching a sensitivity 
limit on the stacked images close to Vega magnitudes $Y$=21.1 mag, $J$=21.3 mag and $K_\mathrm{s}$=20.7 mag 
with a signal-to-noise ratio S/N=10.
The survey covers the LMC area (116 deg$^2$) with 68 tiles\footnote{The ensemble of 16 non-contiguous images 
of the sky produced by an observation with the VISTA IR camera \citep{Eme10} is called ``pawprint''. A ``tile'' is instead a filled area of the sky fully sampled by 
combining six offset pawprints.}, 
while 27 tiles cover the SMC
(45 deg$^2$), and 13 cover the Bridge (20 deg$^2$).
Additionally, 2 tiles (3 deg$^2$) are positioned on
the Stream.
The VMC $K_\mathrm{s}$-band data  are taken over 12 separate epochs (Paper I),   
each reaching a limiting magnitude of 
$K_\mathrm{s} \sim 19.3$ mag with an S/N$\sim$ 5 (Fig.~\ref{fig:errors}).
The  limit reached by the single epoch data allows to 
comfortably detect the minimum light of the RR Lyrae stars in both the LMC and 
the SMC\footnote{Typical average luminosities of the
RR Lyrae stars are $\langle K_\mathrm{s} \rangle \sim 18$ mag and 18.2 mag in the LMC 
and SMC, respectively.}. 
Previous near-infrared surveys such as 
DENIS \citep{Cio00}, 2MASS \citep{Skr06} and  IRSF/SIRIUS \citep{kato07,Ita09} 
were generally single epoch and in any case much shallower than VMC. 
For bright stars, the VMC survey is limited by saturation which causes a 
significant departure from linearity starting at $K_\mathrm{s} \sim$  10.5 mag, the actual value varies with seeing, airmass, etc. 
(see Fig. 7 of  Paper I). As a consequence, the Cepheids for which VMC data are available have pulsation periods shorter than 20 - 30 days. 
The $K_\mathrm{s}$-band monitoring sequence of each VMC   tile can start at any time, but once a sequence is started each subsequent observation
 is obtained at intervals equal to or larger
than: 1, 3, 5 and 7 days, for epochs from 2 to 5, and then, for  epochs from  6 to 11, at
least 17 days apart. 
 This scheduling permits good light curve  coverage for both Cepheids and RR Lyrae stars.

The VMC images are processed by the Cambridge Astronomical Survey Unit (CASU) through the VISTA Data Flow System (VDFS) pipeline that performs aperture photometry 
of the images and computes the  Julian Day (JD) of observation  for each source by averaging only the JDs of the pawprints in which that source was
observed. The reduced data
are then sent to the Wide Field Astronomy Unit (WFAU) in Edinburgh  where the single epochs are stacked and catalogued by
the Vista Science Archive (VSA; \citealt{lewis10, Cro12}). 

\linespread{1}
\linespread{1.3}

\begin{figure}
\includegraphics[width=8cm, height=8cm]{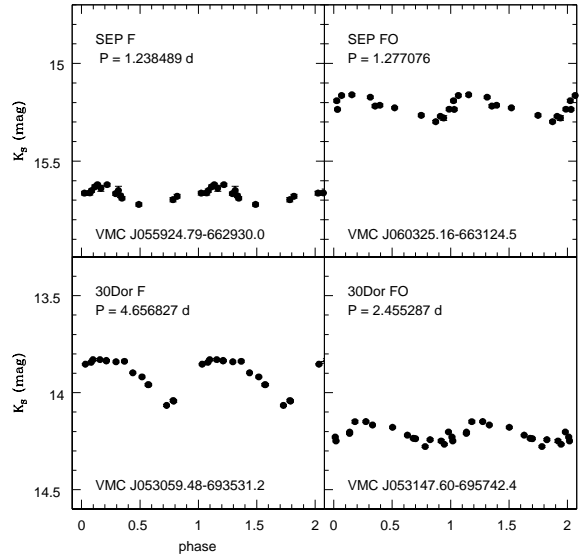}
\caption{Typical VMC $K_\mathrm{s}$-band light curves of a fundamental-mode (F) and a first-overtone 
  mode  (FO) CCs in the 30 
Dor region (lower panels) and in the SEP field (upper panels). 
Errors on the individual $ K_\mathrm{s}$ data-points are as in Fig.~\ref{fig:errors}.
}
\label{fig:lc2-CC}
\end{figure}

\begin{figure}
\includegraphics[width=8cm, height=8cm]{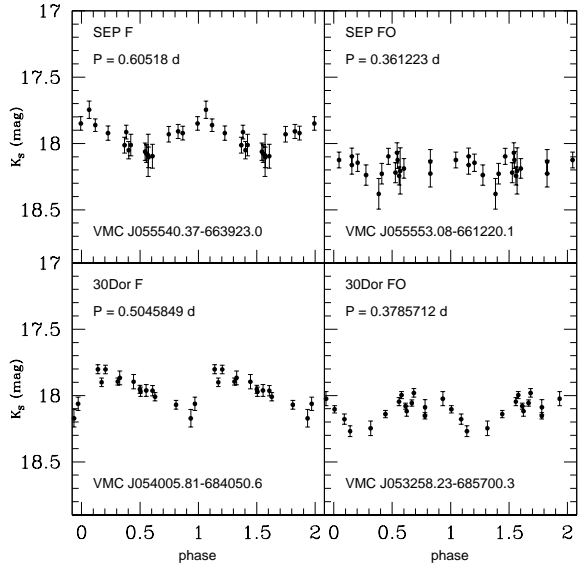}
\caption{Typical VMC $K_\mathrm{s}$-band light curves of fundamental-mode (F) 
 and first-overtone (FO) RR Lyrae stars  in the 
30 Dor region (bottom panels) and 
in the SEP field (upper panels). 
}\label{fig:lc2-RR}
\end{figure}

Full 12-epoch time series were obtained for nineteen  VMC tiles,  as of July 2013: 10 in the LMC, 4 in the SMC, 3 in 
the Bridge, and 2 in the Stream. 
Fig. ~\ref{fig:errors} shows the typical errors of the $K_\mathrm{s}$-band single-epoch data from the VSA aperture photometry for 
CCs (black and cyan points) and RR Lyrae
stars (grey and blue points) in the VMC tiles LMC 6\_6 and 8\_3, respectively.
 The latter tile lies in a low crowding, 
peripheral area of the LMC; tile LMC 6\_6 is centred instead on the well-known 30 
Dor star-forming region where the high crowding boosts up the photometric errors in comparison with the outer fields. 
Two distinct sequences are visible among RR Lyrae stars in tile LMC 8\_3 (blue points,
  see zoomed box). The upper one  corresponds to concatenation mode (shorter observing time, see
  Paper I) observations. 
To this sequence also belong observations obtained on the 25/10/10
and the 22/12/09; 
during the 25/10/10 night the seeing was lower than in other nights. 
During the 22/12/09 night there is no particular issue, apart for a Zero Point smaller respect to other nights. 
RR Lyrae stars in tile LMC 6\_6 (grey points) also show the same dichotomy,
even if it is less evident in Fig.~\ref{fig:errors}; moreover an upper
thin sequence is visible.
This corresponds to observations obtained on the 17/11/2009 when the Zero Point was also lower than in other nights.
Among the sparse fields is tile LMC 8\_8 that includes the Gaia SEP region, 
an area of about 1 square degree in size, that the 
Gaia astrometric satellite (\citealt{Lin96}; \citealt{Lin10}) will 
repeatedly observe for calibration purposes at the start of the mission.
Examples of the VMC $K_\mathrm{s}$-band light curves for CCs and RR Lyrae stars in the 30 
Dor region (lower panel) and in the SEP field (upper panel) are shown in 
Fig.~\ref{fig:lc2-CC} and   Fig.~\ref{fig:lc2-RR}, respectively.
The severe crowding conditions make
the analysis of the RR Lyrae stars in the 30 Dor field 
much more complicated and time consuming than in the SEP field and any other peripheral field 
of the LMC. 
For these stars we used light curves obtained from the PSF photometry technique  
\citep{Rub12} applied on the single
 epoch data as described in Moretti et al. (in preparation). 
In general, we will use 
the aperture photometry processed through the  VISTA pipeline
for the RR Lyrae stars in the outer tiles, and a homogenized 
PSF photometry (see details in Rubele et al.,  in preparation) 
for the highly crowded tiles.
The small error bars in the bottom panels show that the PSF photometry 
is very effective in this crowded field. 
 Moreover, the errors in the upper pannel are smaller or larger
according to the observing conditions (see Fig.~\ref{fig:errors}). 
 
The time sampling of the VMC survey along with the significantly 
reduced amplitude of the light variation in the 
$K_\mathrm{s}$ passband allow us to  obtain very precise estimates of
 the  mean $K_\mathrm{s}$ magnitude of the  MS pulsating variable
 stars (R12b, R13) 
but we have to rely on variable star catalogues produced by the  
microlensing surveys 
(Sections~\ref{sec:microlensingSurvey}) for 
 identification, coordinates and pulsation properties (period, epoch of maximum light, 
parameters of the Fourier decomposition of the visual light curves).  
Unfortunately, parts of the VMC survey's footprint  is not presently covered by the microlensing surveys, leaving us to rely
on the VMC data alone for the identification. 
The average magnitudes in
the $K_\mathrm{s}$-band, derived using a spline interpolation of the data for the CCs and the fit with templates (\citealt{Jon96}) for the RR Lyrae stars, are then 
used to construct $PL$, $PLC$ and $PW$  relations that, thanks to
their small intrinsic dispersions, can provide individual distances to the
investigated variable stars and in turn information on the 3-D structure of
the MS.

\subsection{Optical catalogues:  the microlensing surveys}\label{sec:microlensingSurvey}
\begin{figure*}
\begin{centering}
\includegraphics[width=8.5cm, height=8.5cm]{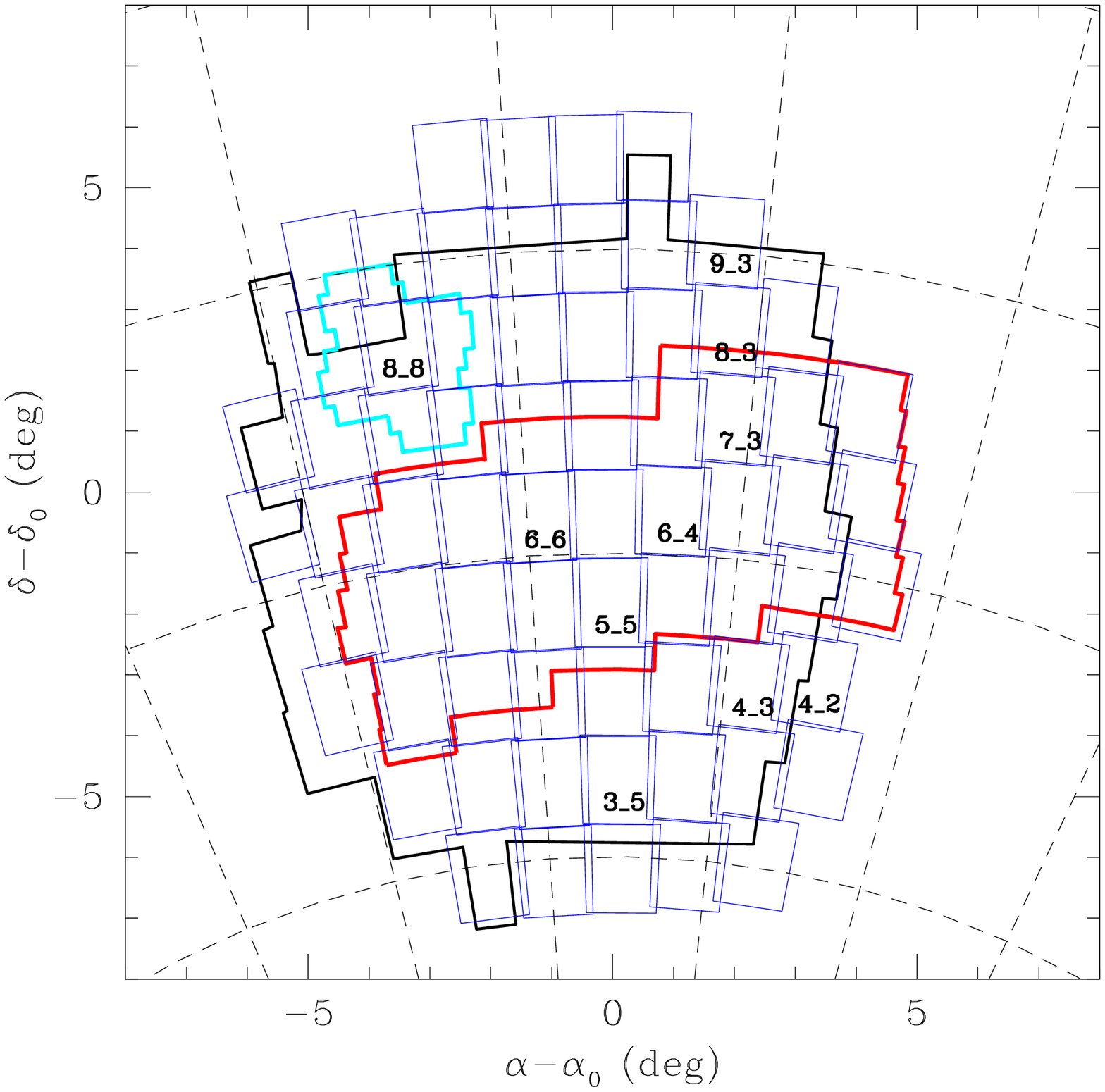}
\includegraphics[width=8.5cm, height=8.5cm]{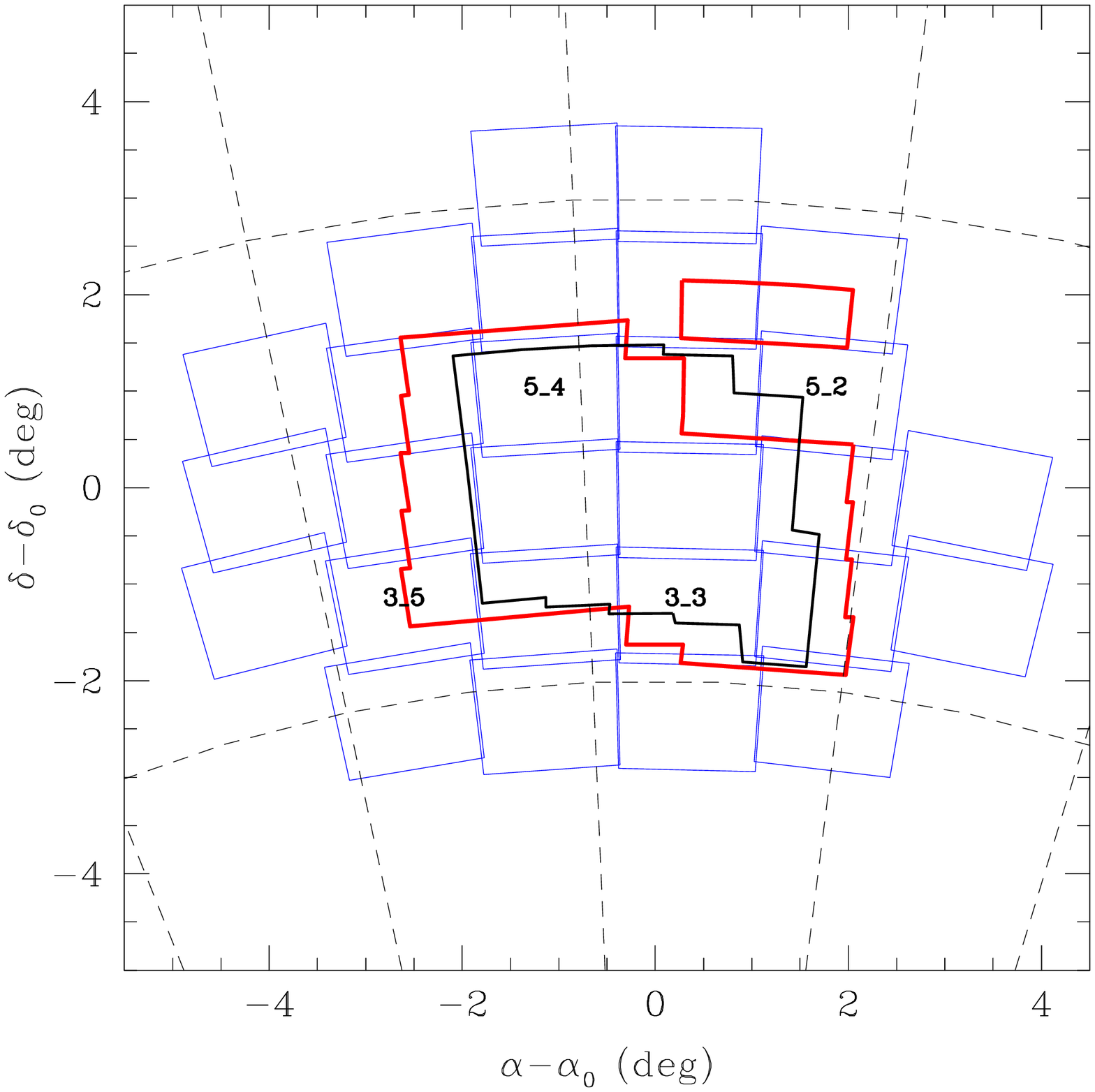}
\includegraphics[width=8.5cm, height=8.5cm]{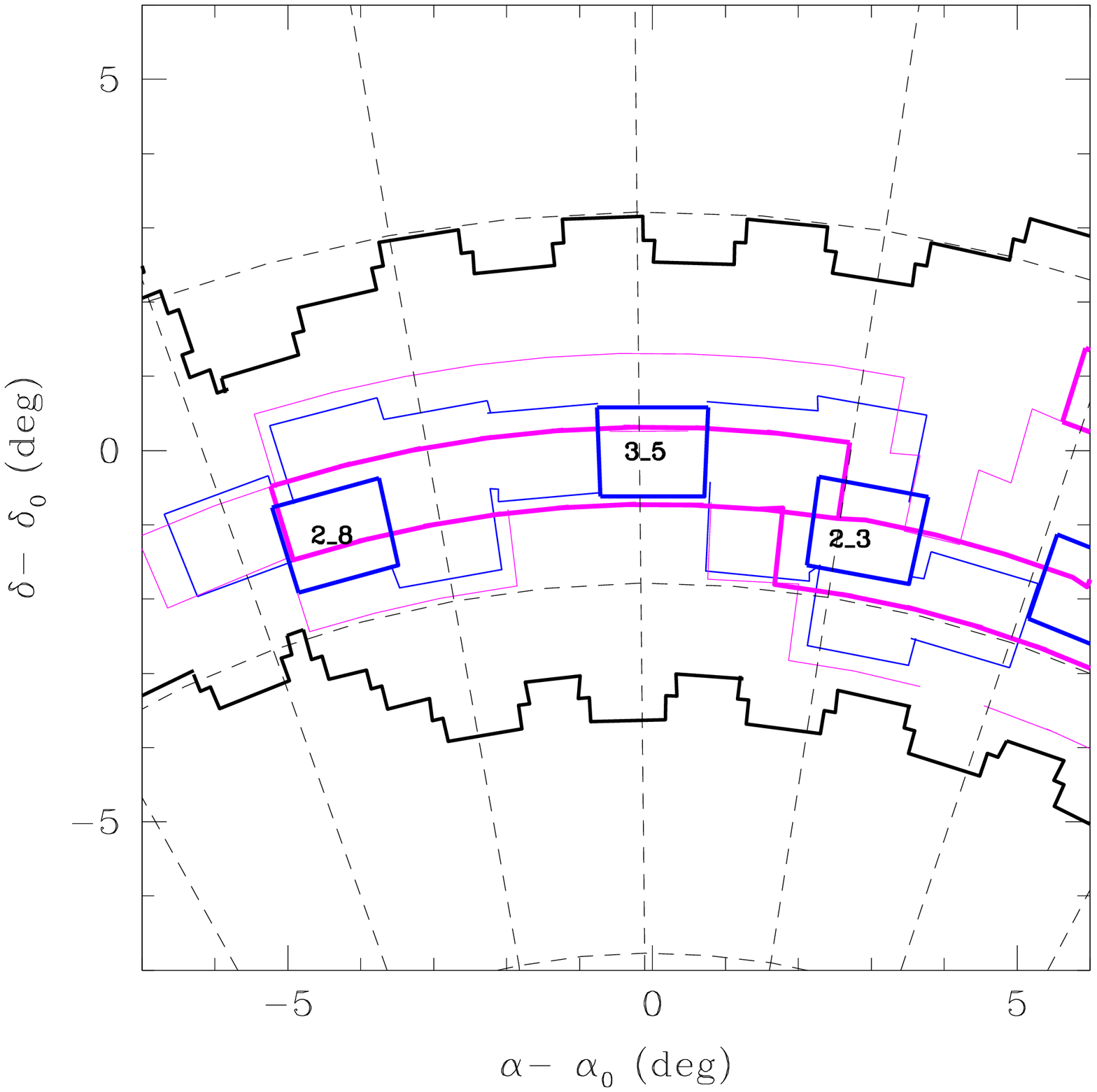}
\caption{Upper panels: Sky coverage of the LMC (left) and SMC (right) 
 for  EROS-2 (black),  OGLE~III (red)  and OGLE~IV (cyan). VMC tiles are indicated in blue; tiles already completely
 observed are labelled.
Lower panel: Map of the Bridge area. The sky coverage of OGLE~IV (black), VMC (blue) and STEP (magenta) are shown. 
Thicker lines show the VMC and STEP areas completely observed as of July 2013 in
both surveys; VMC tiles already completely
 observed in the Bridge area are labelled.}
\label{fig:ogleObs} 
\end{centering}
\end{figure*}

The LMC/SMC microlensing surveys have  detected and characterized 
tens of thousands  
RR Lyrae stars, CCs, binaries,  and
Long Period Variables (LPVs).  
So far, the largest spatial coverage of the LMC is that obtained by the second 
generation of the EROS microlensing experiment (hereinafter EROS-2, \citealt{Tis07}).  
 EROS-2 time-series data were collected in two passbands, 
 a  $blue$ channel with $\lambda =$  (420 - 720) nm, that overlaps to the $V$ 
and $R$ standard bands, and a  $red$ channel with $\lambda  = $
(620 - 920) nm, 
that roughly matches the mean wavelength
of the Cousins $I$ band \citep{Tis07}.
 OGLE \citep{Uda97}, of which the most extended area coverage is so far that obtained during the 
third phase (hereinafter OGLE~III)  uses instead 
 standard $B_{\rm Johnson}$ ($B_J$), $V_{\rm Johnson}$ ($V_J$) 
and $I_{\rm Cousins}$ ($I_C$) filters. Furthermore,  the median seeing of the OGLE~III images is better than the one of EROS-2 images.
 It allows a better stars separation and measurement in highly crowded fields. 
For these reasons the use of the OGLE~III data was preferred wherever those data were available (e.g., in 
the regions covering the  LMC central bar), while we are using the EROS-2 data in the outer parts of the LMC that 
are not yet covered by  other surveys. 
For the SMC, the areas covered by  EROS-2 and
OGLE~III are very similar, hence we will mainly use as a reference for 
our study the OGLE~III data.
Fig.~\ref{fig:ogleObs} shows the sky
coverage of different surveys in different regions of the MS. The distribution
of VMC (blue boxes), OGLE~III and IV (red and cyan contours, respectively) and 
EROS-2 (black line) field of view (FoV) are shown for the LMC (upper left panel) and the SMC
(upper right panel).
The OGLE~III catalogues of variable stars are publicly available at the web site\footnote{
http://ogle.astrouw.edu.pl/}, 
 and contain 
light curves for 24906 RR Lyrae stars and 3361 CCs in the LMC, and 
2475 RR Lyrae and 4630 CCs in the SMC.
For each object the catalog provides right ascension, declination, 
mean Johnson-Cousins $V,I$ magnitudes, period, $I$-band amplitude, along 
with the Fourier parameters R21, $\phi$21, R31, and $\phi$31 of the $I$-band light curves (\citealt{Sos09}).
On the other hand, the EROS-2 catalogues of RR Lyrae stars and Cepheids are not public yet, but they 
were kindly made available to us 
by the EROS-2 collaboration. 
It should be noted that  OGLE III has published  catalogues of  confirmed 
CCs and RR Lyrae stars for both the LMC and the SMC (\citealt{Sos08a,
Sos09, Sos10a, Sos10b}) whereas for EROS-2 we only have 
catalogues of candidate Cepheids and RR Lyrae stars, 
that we individually checked, as described in Section~\ref{sec:EROS_LMC}.

A number of the most peripheral VMC tiles in both the LMC and the SMC and, most 
noteworthy, the entire Bridge region between the two Clouds, are not covered yet
by any of the previous microlensing surveys. However, optical time-series data will be 
provided both in the SMC and in the Bridge area by the survey 
\emph{``SMC in Time:
Evolution of a Prototype interacting late-type dwarf galaxy''} (STEP, P.I.
V. Ripepi; see \citealt{Rip06}) which is being carried out with the VLT Survey Telescope
VST\footnote{http://vstportal.oacn.inaf.it/}. Furthermore, the OGLE~IV\footnote{see http://ogle.astrouw.edu.pl/} 
survey, started in 2010, is in progress. Once completed it will 
almost entirely cover the whole FoV of the VMC survey. 
The lower panel of Fig.~\ref{fig:ogleObs} 
shows the sky coverage in the Bridge area of 
the OGLE~IV (black contours), VMC (blue boxes) and STEP (magenta boxes) surveys. 
Thicker lines mark tiles already completely observed.  

Finally, no 
 pulsating variables are known in the Stream. It is supposed to be mainly 
gaseous but recently the presence of a stellar Stream  
 counterpart was predicted and suggested (\citealt{Bes13}; \citealt{Bag13}; \citealt{Noe13}.  
This makes it crucial to find a method for selecting variable stars from the VMC data  alone \citep{Cro12}.

\subsection{Combining OGLE, EROS and VMC data for the variable stars}\label{subsec:Combining}

\begin{figure}
\begin{centering}
\includegraphics[width=7cm, height=7cm]{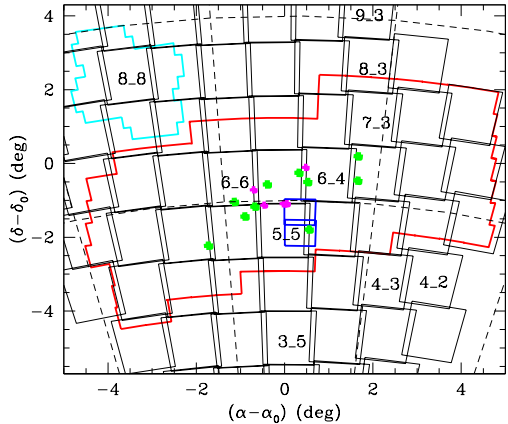}
\caption{Location of the  ten LMC tiles completely observed by VMC as of July 2013 
  with respect to the OGLE~III and OGLE ~IV (red and cyan contours, respectively) fields. 
  Tiles LMC
5\_5, 6\_4, 6\_6 and 7\_3 are entirely covered by the OGLE~III observations.  Tile LMC 8\_8 is entirely covered by OGLE~IV, and also for about 70\% by EROS-2. 
Tile LMC 8\_3  is  entirely covered by EROS-2 and only for about  30\% by OGLE~III. EROS-2 covers 
entirely tiles LMC 3\_5 and 4\_3, roughly 40\% of tile LMC 9\_3, and only  a very tiny portion of tile LMC 4\_2 (see also Fig.~\ref{fig:ogleObs}).
Also marked are the LMC regions where a number of  RR Lyrae stars were studied spectroscopically  
by \protect\cite{Cle03} and \protect\cite{Gra04} (blue rectangles),  
\protect\cite{Bor04,Bor06,Bor09} (green boxes), and  \protect\cite{Sze08} (magenta boxes).  
}
\label{fig:LMCmet}
\end{centering}
\end{figure}

Figs.~\ref{fig:ogleObs} and ~\ref{fig:LMCmet} shows the location of the ten 
tiles  completely observed in the LMC as of July 2013,  with respect to the OGLE~III and the recently published first 
OGLE~IV data. 
The  discussion  in the following Sections will 
refer mainly to the tiles LMC 5\_5, 6\_4, 6\_6, 8\_3, and  8\_8, 
for which  fully processed and catalogued single epoch data are already available 
through the VSA, and, more specifically to the tiles LMC 6\_6  (30 Dor field), 8\_3 and 8\_8 (SEP field).

In order  to test our procedures,  for the tiles LMC 6\_6 and 8\_8 we have used  both the 
catalogues of candidate variables 
provided by  EROS-2 and the catalogues of RR Lyrae stars and Cepheids published 
by OGLE~III and IV.  For the other tiles we have used the OGLE catalogues when available, 
otherwise we have exploited the EROS-2 data. 
We have cross-matched the optical catalogues against the
 VMC deep tile (vmcSource table) catalogues 
 using the VSA utilities. Specifically, we used the VMC
tiles coordinates as in Paper I, the VMC LMC tile dimension of 1.201 degree in RA, 
1.475 degree in DEC, and the angular coordinates as described in \cite{van01} 
to select the optical data, then we used the Cross-ID tool available in the VSA \citep{Cro12} to match 
the optical and the VMC infrared catalogues. 
A small pairing radius of 0.5 arcsec was adopted 
for tiles LMC 5\_5, 6\_4 and 6\_6  to  reduce the number of misidentifications in  these rather crowded regions of the LMC. 
It was increased to 1.0 arcsec  
 for the outer LMC tiles 8\_3, 8\_8, which are less affected by 
crowding. 
 Table~\ref{tab:opt_data} summarizes the numbers of 
CCs and RR Lyrae stars discovered by the microlensing surveys in the tiles discussed here, and the numbers of their VMC counterparts. 
The incomplete optical coverage of some tiles explains the small number of known variables. 

The recovery of infrared counterpart of RR Lyrae stars varies from 74\% in the crowded inner tiles to 98\% in the outer ones.
The recovery rates for the intrinsically brighter CCs are higher:  98\% in the central tiles and 100\% in the outer ones. 
Tests performed on tile LMC 6\_6 show that the lower completeness of the internal 
fields is mainly due to Cepheids not recovered in the VSA VMC catalogue because are at the edges of the tile. 
In fact,  even in this most crowded tile
the completeness rises to 100\% in the internal part of the tile.
The lower completeness for the RR Lyrae stars is instead mainly due to the 
high crowding in the internal regions  and nebular gas emissions which limit the
detection of variables 
as faint as the RR Lyrae stars.  In these conditions the PSF photometry shows more accuracy and efficiency than the
VSA aperture photometry, 
increasing the number of RR Lyrae stars with VMC $K_\mathrm{s}$ photometry. Indeed, in 
the central part of tile LMC 6\_6 the completeness of the VSA Catalogue 
for RR Lyrae stars is 82 \%, but it rises to  
 86\% for the PSF catalogue. Increasing the pairing radius from 0.5 to 1 arcsec, would also increase the number of cross-matches 
 up to about 90\%.  However,  the comparison of optical and infrared light curves, and the
position in the ($K_\mathrm{s}$, $I$-$K_\mathrm{s}$) CMD reveal that 47 of 51 additional detections obtained using a pairing radius
between 0.5 and 1 arcsec, are not true RR Lyrae stars. 
 Therefore using 
 a pairing radius of 0.5 arcsec in
the crowded regions is preferred because it yields a  more reliable sample.

\begin{table*}
\caption{Statistics of RR Lyrae stars and CCs within the VMC tiles: each column lists the number 
of variables found by the microlensing surveys, the number of the 
recovered VMC counterparts, and the fraction of variables with VMC counterparts. The 
paring radii were 0.5 arcsec for 
the inner LMC regions, and 1 arcsec for the outer regions (see Section~\ref{subsec:Combining} for details).}
\label{tab:opt_data}
\begin{tabular}{|c|c|c|c|c|cl} 
\hline
Tile &       \multicolumn{2}{c}{OGLE~III}                 &     \multicolumn{2}{c}{EROS-2} &  \multicolumn{2}{c}{OGLE~IV} \\
        &              RRL  &  CC &   RRL & CC & RRL & ~~~~CC\\
\hline
\hline
5\_5 &	    2753 (2255) 82\% &    214 (207) ~97\%&  $-$~~ $-$~~ $-$&  $-$~~ $-$~~ $-$ & $-$~~ $-$~~ $-$&$-$~~ $-$~~ $-$\\
6\_4 &	    3446 (2543) 74\% &    402 (393) ~98\%&  $-$~~ $-$~~ $-$&  $-$~~ $-$~~ $-$ & $-$~~ $-$~~ $-$&$-$~~ $-$~~ $-$\\
6\_6 &           2040 (1637) 80\% &    327 (321) ~98\%&  $-$~~ $-$~~ $-$&  $-$~~ $-$~~ $-$ & $-$~~ $-$~~ $-$&$-$~~ $-$~~ $-$\\
\hline
8\_3 &	 ~~133$^{a}$ ~(127)  95\% &  ~52$^{a}$ ~(52) 100\%&  262~~ (258) 98\%& 126 (125) ~99\% &  $-$~~ $-$~~ $-$&$-$~~ $-$~~ $-$\\
8\_8 &            $-$~~ $-$~~ $-$    &     $-$~~ $-$~~ $-$&  109$^{b}$ (109) 100\%&~~9$^{b}$ ~~(9) 100\%  &223 (219) 98\%   &19 (19) 100\%\\
\hline
\end{tabular}
\\
$^{a}$ \footnotesize{OGLE~III covers only a tiny portion corresponding to about 1/4 of tile LMC 8\_3.}\\
$^{b}$  \footnotesize{EROS-2 covers only about 2/3 of tile LMC 8\_8.}\\
\end{table*}

\section{The central fields of the LMC}\label{sec:ogle30Dor}

A number of different studies, besides the microlensing surveys,  have targeted the RR Lyrae stars in the central 
region of the LMC both photometrically and spectroscopically.
\cite{Cle03} and \cite{Gra04} presented visual photometry and spectroscopy
for more than a hundred RR Lyrae stars in two fields located close
to the bar of the LMC (blue rectangles in Fig.~\ref{fig:LMCmet}).
They measured average magnitudes, local reddening, and individual metallicities for the RR Lyrae stars. 
In particular,  they inferred  a mean metallicity for the RR Lyrae stars in this region 
of the LMC of [Fe/H]$=-1.48\pm0.03$ dex,  with $\sigma=0.29$ dex on the
Harris (1996) metallicity scale.  Tile LMC 5\_5 cointains both fields observed by \cite{Cle03} and \cite{Gra04}.
We will use the metallicities in those papers to study the $PL_KZ$ relation of the RR Lyrae stars in tile LMC 5$\_5$. 
\cite{Bor04,Bor06} measured radial velocities for 87 LMC RR Lyrae stars and 
metallicities
for 78 of them. These targets are located in 10  fields (green rectangles
in Fig.~\ref{fig:LMCmet}) spanning a wide
range of distances, out to 2.5 degrees from the center of the LMC. One of these 
fields is contained in the 30 Dor tile.
They inferred a mean metallicity value [Fe/H]=$-1.53\pm0.02$ 
dex  with a dispersion of $\sigma=0.20\pm0.02$ dex using the
\citet{Gra04} technique, and showing 
that the RR Lyrae stars in the central part of the LMC form a rather homogeneous
metal-poor population. We will use the metallicity values obtained by Borrisova et al. 
 for the RR Lyrae stars in the 30 Dor field (Moretti et al. in preparation). 
For the RR Lyrae stars in the tiles for which there are no spectroscopic measurements, metal abundances will be 
estimated from the Fourier parameters of the $V$-band light curves according to the technique devised by \cite{JK96}  and  \cite{Mor07} for 
fundamental-mode and first-overtone RR Lyrae, respectively, and adopting the new metallicity calibrations  obtained by 
\cite{Nem13}. 

Near-infrared $PL$ relations for the RR Lyrae stars in the central part of the 
LMC have been obtained by a number of different  authors.
 For example, \cite{Sze08} obtained deep near-infrared $J$ and $K$ band
observations of six fields, three of which overlap,
located in the LMC bar (magenta rectangles in Fig.~\ref{fig:LMCmet}). 
They found consistent values for the distance modulus of the LMC  using a number of  
different theoretical and empirical calibrations of the $PL_KZ$
relation, and adopt as their final value  $18.58\pm0.03 \rm (statisical) \pm (0.11)
\rm (systematic)$ mag, in good agreement with most independent determinations of the distance to 
this galaxy.
 \cite{Bor09} investigated the metallicity dependence of the near-infrared $PL$ 
relations for RR Lyrae stars 
combining near-IR photometry and \cite{Bor04,Bor06} spectroscopically measured metallicities for 
50 RR Lyrae stars. 
They  found a very mild dependence on metallicity of the $PL$ in the $K$ band,  and 
inferred from  their near-IR $PL_KZ$ 
relation an LMC  distance modulus of $18.53\pm0.13$ mag. They point out  
that their distance modulus
relies on the trigonometric parallax of RR Lyrae itself. We will use 
these works as a reference in our studies of the near-infrared $PLZ$ relation for the RR Lyrae
stars that will be described in subsequent papers.

\begin{figure*}
\includegraphics[width=8cm, height=8cm]{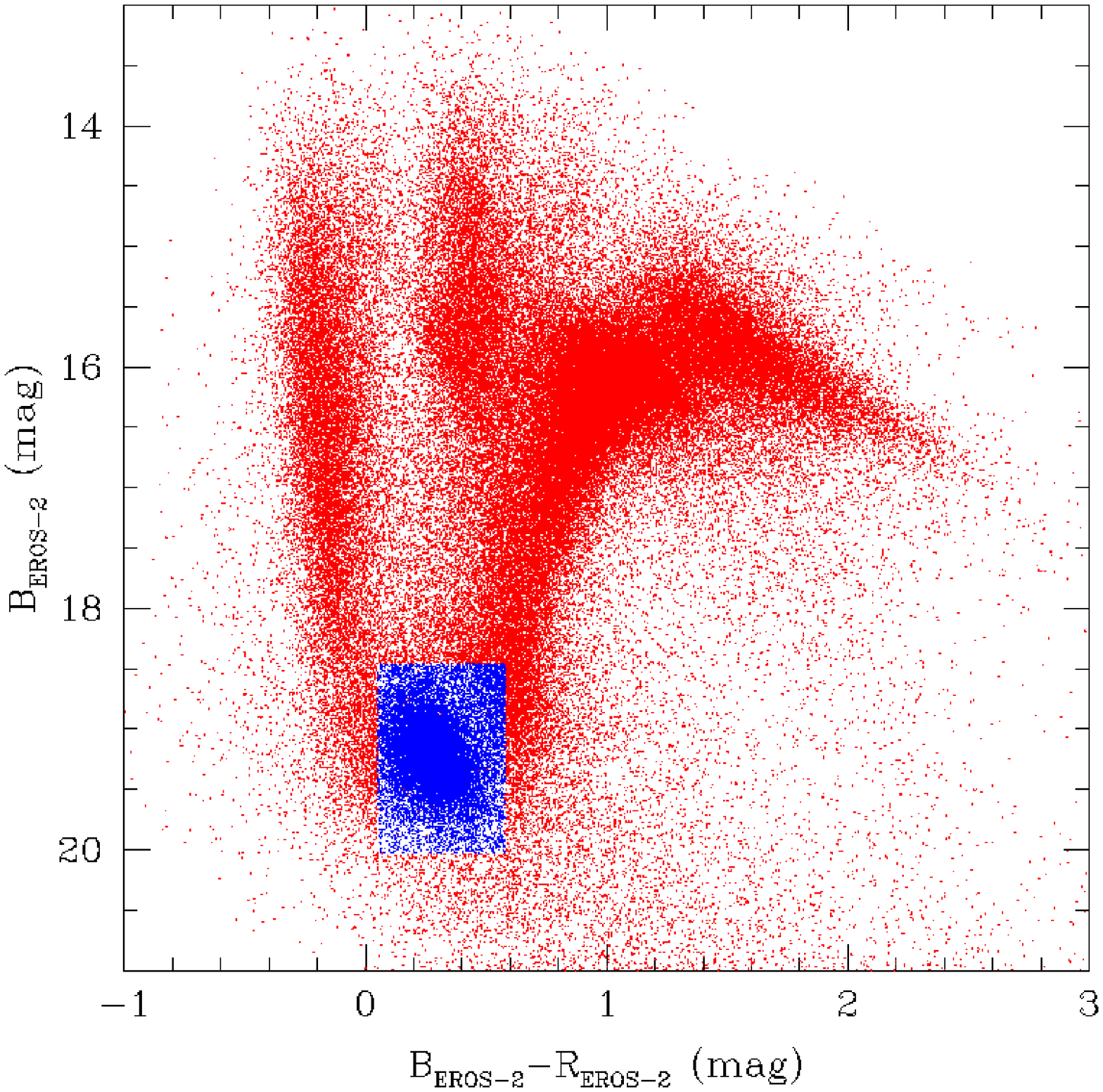}
\includegraphics[width=8cm, height=8cm]{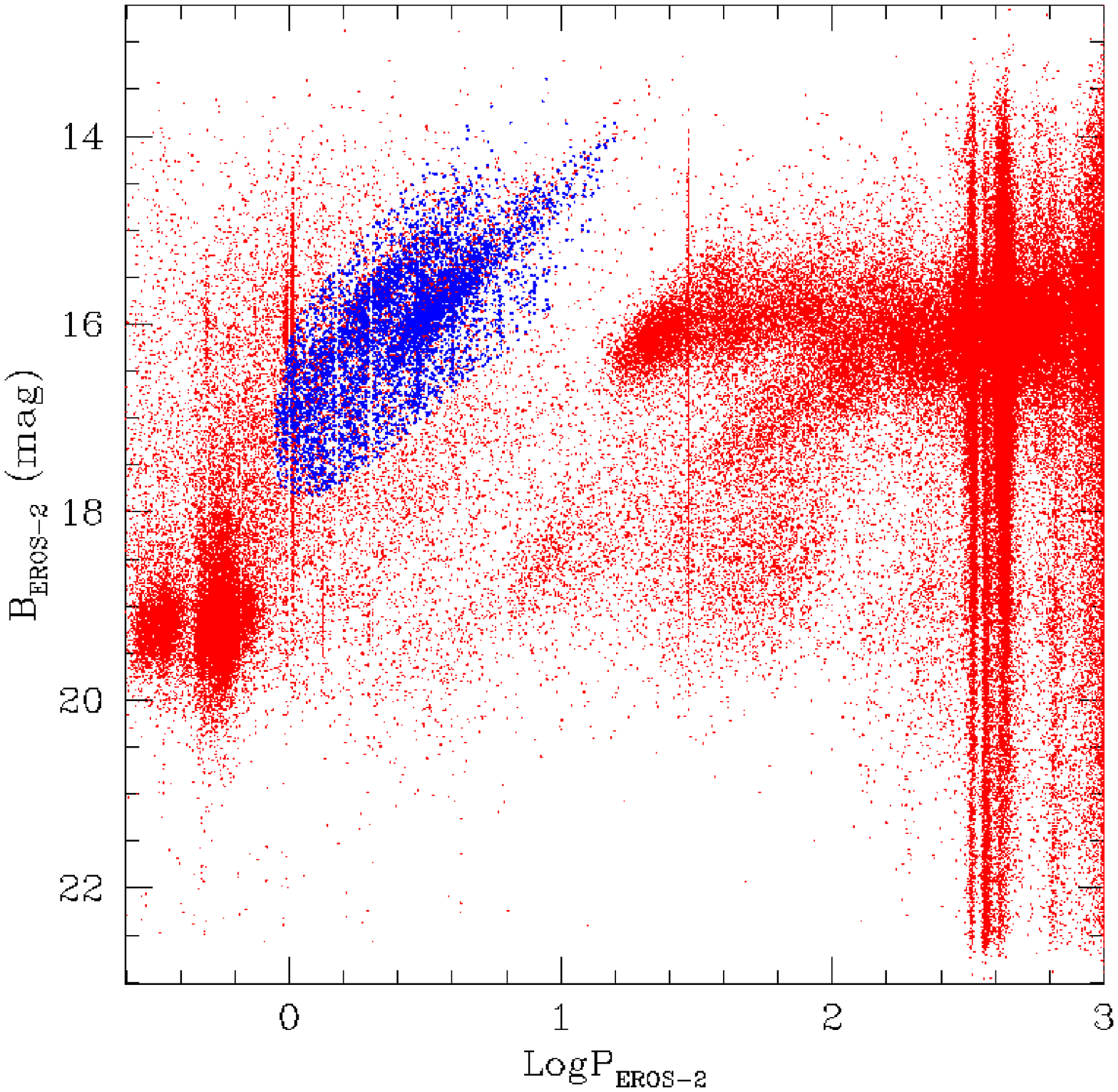}
\caption{Left panel: $B_{EROS}$, $B_{EROS}-R_{EROS}$ CMD of LMC candidate variable stars from the EROS-2
data (red points). 
A blue box marks the region populated by the RR Lyrae candidates; \label{fig:LMCselrr}  
Right panel: Distribution of  the EROS-2 candidate variables in the LMC 
(red points) in the LogP, $B_{EROS}$ plane. 
Blue points mark the candidate 
CCs. \label{fig:LMCselcc}}
\end{figure*}

\section{The outer fields of the LMC}\label{sec:EROS_LMC}
For the  study of the variable stars in the outer  fields of the LMC, which were not covered by OGLE~III  and for which 
the OGLE~IV data are not available yet we are extensively using the EROS-2 data, kindly provided by the EROS collaboration 
in the form of lists of
``candidate'' RR Lyrae stars and Cepheids, separately.  
The detection of variable stars in EROS-2 was 
performed by an automatic pipeline based on the Analysis of Variance (AoV) method and 
software developed by \cite{Bea97} and \cite{SC03}:  
stars  with AoV statistics $\geq$ 20 were considered candidate variables (see \citealt{Mar09}
and references therein for further details).
The left panel of 
Fig.~\ref{fig:LMCselrr} shows the ($B_{EROS}$, $B_{EROS}-R_{EROS}$) colour
magnitude diagram (CMD) obtained from the EROS-2 catalogue of the LMC candidate variables (red
points). The candidate RR
Lyrae stars were extracted 
by selecting in this CMD 
objects with the
$18.46 < B_{EROS} < 20.03$ mag,  and $0.05 < B_{EROS}-R_{EROS} < 0.58$ mag (blue points). 
The selection of the CCs was based instead on the EROS-2
Period versus $B_{EROS}$-magnitude diagram (see right panel of Fig.~\ref{fig:LMCselcc}) 
by considering objects with  $\sim 13.39 < B_{EROS} < \sim 17.82$ mag, and 
 $\sim 0.89 < P < \sim 15.85 $ days for the period (blue points).

These selection criteria returned a  list of 16337 candidate RR Lyrae stars 
 and 5800 candidate Cepheids
over the whole sky area covered by EROS-2 in the LMC.

\subsection{Tile LMC 8\_8: EROS-2 data} 
\label{subsec:gratis}
\begin{figure}
\begin{centering}
\includegraphics[width=8.5cm, height=8.5cm]{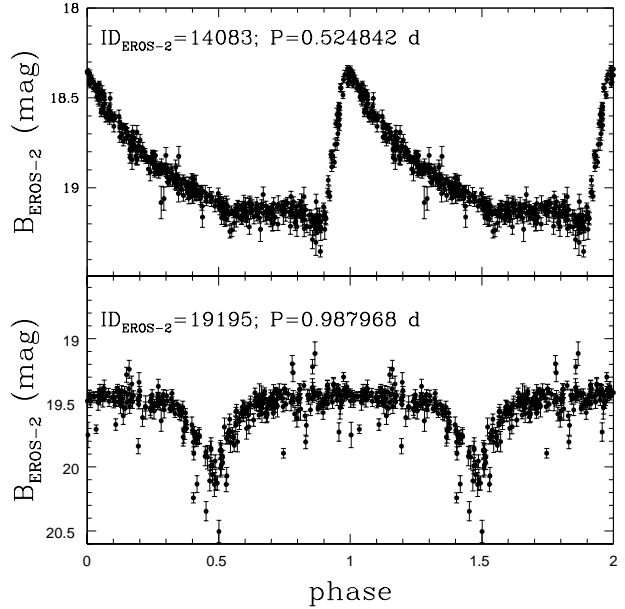}
\caption{Upper panel: $B_{EROS}$-band light curve of a confirmed 
fundamental mode RR Lyrae star in the SEP;  Lower panel:
Light curve of a binary system found among the sample of EROS-2 candidate 
RR Lyrae stars in the SEP field.
\label{fig:lc}}
\end{centering}
\end{figure}

The EROS-2 candidate RR Lyrae stars and Cepheids in the SEP field were
extracted by selecting from the EROS-2 catalogue objects  that lie in the
corresponding region (see Appendix for details). 
The selection returned a
list of 14  CC candidates, all of which have a VMC counterpart, and 123 candidate RR Lyrae stars of which 122 have a VMC counterpart
within 1 arcsec\footnote{The only missing star has a possible VMC counterpart at
distance $\sim$1.7 arcsec, which  appears to be a  blend after visual inspection of the
VMC images.}.
To check classifications and periods we studied the optical light curves of these objects with the 
GRaphical Analyser of TIme Series (GRATIS) package,
custom software developed at the Bologna Observatory by P. Montegriffo 
(see e.g., \citealt{Cle00}). The visual inspection of the light curves allowed us to clean the samples of candidate RR 
Lyrae stars  and Cepheids from spurious sources 
 and to identify wrong periods, as for instance in the case of star 
ID$_{EROS2}$=lm0383l16657, for which P$_{EROS-2}$ is twice the actual  period. 
Fig.~\ref{fig:lc} shows the light curve of two sources classified as RR Lyrae 
candidates according to the selection criterion described above:  
a confirmed RR Lyrae star on the top, and 
  a contaminant binary on the bottom. 
We  checked the EROS-2 period (P$_{EROS-2}$) and obtained our own period estimates (P$_{GRATIS}$) 
for all the EROS-2 candidate Cepheids and for a subsample of 24 RR Lyrae stars 
in the SEP field.
We generally found a very good agreement between P$_{EROS-2}$ and P$_{GRATIS}$, 
with difference, on average, on the fifth 
decimal place,  thus we have  adopted P$_{EROS-2}$ for the remaining stars.
Although we did not calculate 
P$_{GRATIS}$ for all candidate RR Lyrae stars and Cepheids in the SEP field,  we visually inspected the light curves
of all of them using the GRATIS package and the P$_{EROS-2}$ as input.
With GRATIS, we also determined the epoch of maximum light
for each $B_{EROS}$-band light curve. This information was later used
along with the period to fold the $K_\mathrm{s}$-band light curves as
described in R12a.
The visual inspection of the  light curves for 14 EROS-2 CC candidates in the SEP area, confirmed 9 CCs: one fundamental-mode (F), 7 first-overtone mode 
(FO), and one double-mode Cepheid, in agreement with results published in \cite{Mar09}; 
 3 turned out to be eclipsing binaries (ECLs),  and
 the remaining two were found to  have very small amplitudes (of the
 order of about 0.1 mag in the EROS-2 bands and of 0.02-0.03 mag in
 the $K_\mathrm{s}$ band).   
Period and luminosity place these two stars in the Cepheid's domain, but their 
 classification is uncertain (see Section~\ref{subsec:ogleiv}).  Final classification and period for the candidate Cepheids in the SEP tile are summarized in  
Table~\ref{tab:infoCC_GSEP}. 
On the other hand, the analysis with GRATIS  revealed that the sample 
of 123 EROS-2 candidate RR Lyrae stars in the SEP field
contained (see
 Table~\ref{tab:infoRR_OGLEIV_GSEP}) 79 
fundamental-mode (RRab), 23 first-overtone (RRc), 7 double-mode (RRd) RR Lyrae stars, 4 ECLs,
2 short period (``s.p." in 
Table~\ref{tab:infoRR_OGLEIV_GSEP})  
variables with light curves similar to RR Lyrae stars but having shorter 
periods and bluer colours,  one object with  singular shape of the light curve 
(``s.s.'' in 
 Table~\ref{tab:infoRR_OGLEIV_GSEP}),  and 7 variables 
that according to EROS-2 have periods between 37 and 1103 days. The latter  
appear to  have very noisy and often irregular light curves with $B_{EROS}$ 
within 19 - 20 mag, and period aliases of  0.33, 0.50, and 0.99
days. 
Some of these objects may indeed be long period variables, we flagged them 
as non classified (``n.c.") objects in 
Table~\ref{tab:infoRR_OGLEIV_GSEP}.

\begin{figure}
\begin{center}
\includegraphics[width=8.5cm, height=8.5cm]{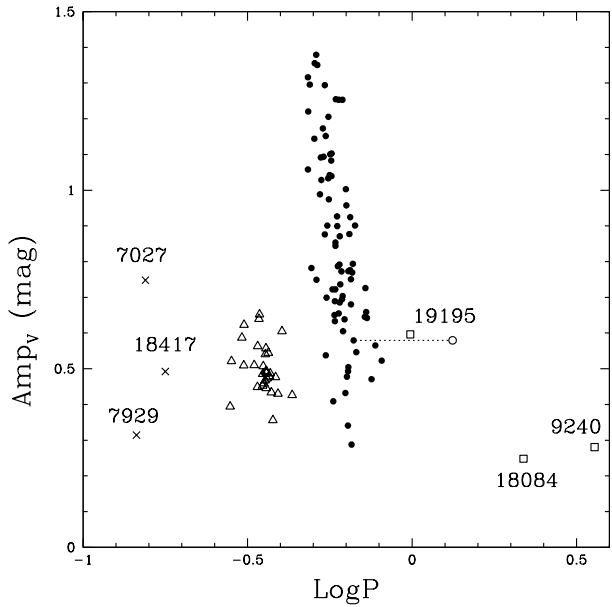}
\end{center}
\caption{Period-amplitude diagram in the $V$ band of the EROS-2 candidate RR Lyrae stars in the SEP field. Objects deviating from the RRab (filled circles) and RRc (open triangles) distributions are short period variables (crosses), and 
ECLs (squares). An open circle marks an RRab star for which EROS-2 period is an alias. The star perfectly falls on the RRab distribution when the correct
 period, half of the EROS-2 value, is adopted. 
 }
 \label{fig:bailey}
\end{figure}

Additional tools  we can use to classify the RR Lyrae stars,  to identify their  pulsation mode, and to infer their metal abundance 
are the period amplitude diagram (LogP, A$_V$; Bailey diagram, \citealt{Bai1902}) and the Fourier parameters of the light curve 
decomposition.
 Since light curves in the Johnson $V$ band ($V_J$) are needed to be able to use these tools we 
 transformed the $B_{EROS}$ and $R_{EROS}$ light curves 
 of the RR Lyrae star candidates to the  
$V_J$ magnitude using Eq. 4 in \cite{Tis07}:
\begin{equation}\label{eq:conversionR}
R_{EROS} = I_C;   
\end{equation}
\begin{equation}\label{eq:conversionB}
B_{EROS}=V_J-0.4(V_J-I_C).
\end{equation}
and then analyzed the $V_J$  light curves with GRATIS.
This was necessary to check the quality of the $V_J$  light curves
obtained with the Eqs.~\ref{eq:conversionR} and \ref{eq:conversionB} and also 
to infer the $V$ amplitude ($A_V$) which is 
needed  both to classify  the RR Lyrae stars in types with the period-amplitude diagram   
and to fit  the $K_\mathrm{s}$ band light
curve using the template fitting method (\citealt{Jon96}, see also
discussion in R12a). 
Fig.~\ref{fig:bailey} shows the LogP, A$_V$ diagram 
of the EROS-2  candidate RR Lyrae stars in the SEP field. We have not plotted sources with extraordinarily long periods (P$>$ 10 days;  see 
Table~\ref{tab:infoRR_OGLEIV_GSEP})
that accidentally fall into the RR Lyrae sample.
In this plot we can distinguish the following groups: 
\begin{itemize}
\item Log$P < -0.6$ and   Log$P > 0.2$ - the analysis of the light
  curves revealed that the former are short period variables (black crosses; flags s.p. and s.s. in Table~\ref{tab:infoRR_OGLEIV_GSEP})  and the latter are ECLs 
(empty squares, ECL in Table~\ref{tab:infoRR_OGLEIV_GSEP});
\item $-0.6 \leq$ Log$P \leq -0.3$ and $A_V \leq 0.7$ mag - these are
  first-overtone ($RR_c$)  RR Lyrae  stars (empty triangles);
\item $-0.3 \simeq$ Log$P \simeq 0.1 $ - these are fundamental mode
  ($RR_{ab}$) RR Lyrae stars (filled points).
\end{itemize}

The classification based on the period-diagram fully confirms the classification 
obtained through the visual inspection of the light curves, and shows that this method can be successfully used to obtain a first 
check and classification of the EROS-2 candidate RR Lyrae stars, that can then be refined by visual inspection of the objects most deviating
from the main distributions of RRc and RRab stars in the Bailey diagram. This is, in fact, the procedure we will adopt  to analyze the sample of EROS-2 candidate RR Lyrae stars
in the tiles covered only by EROS-2. 
Only two objects (namely, idEROS-2: lm0385k3074 $P \sim 0.40$ days, and lm0507k.19195 $P \sim 0.99$ days) 
are classified  respectively as RRc and RRab  by  the LogP, A$_V$ 
diagram, but were found to be binary systems by the visual inspection of the light curves. This corresponds to a $\sim$ 2\% residual contamination.

\begin{figure}
\begin{center}
\includegraphics[width=8.5cm,height=8.5cm]{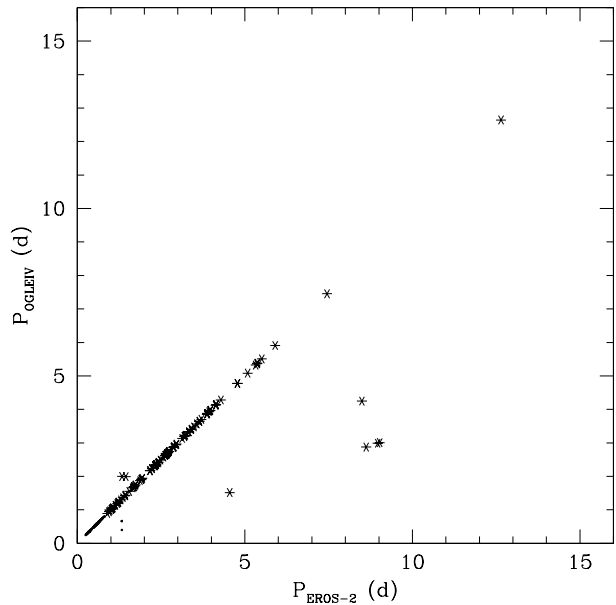}
\caption{Comparison between P$_{EROS-2}$ and P$_{OGLE~IV}$ for RR Lyrae  
stars (points) and CCs (asterisk)  in the OGLE~IV GSEP area in common between the two 
surveys (see text for details). Note that the errors on the periods are
smaller than the point dimensions.
}
\label{fig:checkP_EROS-2_OGLEIV}
\end{center}
\end{figure}

To summarize, the final catalogue of bona-fide RR Lyrae stars identified  in  the portion of the SEP tile  covered by EROS-2 includes 
79 RRab, 23 RRc, and 7 RRd variables for a total number 
of 109 confirmed RR Lyrae stars.

\subsection{Tile LMC 8\_8:  Comparison of the OGLE~IV and EROS-2 data}\label{subsec:ogleiv}
OGLE~IV data for the variables in the SEP field were published by 
the OGLE team while we were writing  this paper, 
thus making it possible for us to include them in our analysis. 
The so called Gaia SEP (GSEP) field (cyan contours in
Figs.~\ref{fig:ogleObs} and ~\ref{fig:LMCmet}; \citealt{Sos12})  
is covered by four OGLE~IV pointings 
 and extends over a total area of about $\sim$ 5.3 deg$^2$, of which the central one square degree corresponds 
to the region that Gaia will repeatedly observe during commissioning. 
The OGLE collaboration opted  to 
make the GSEP field
data available after only two years of observation because of the potential these data 
can have for 
 the Gaia mission. The dataset consists of  $V$ and $I$ bands photometry 
for 6789 variable stars,  with a number of data points between 338 and 351 in $I$,  and about  29 epochs in $V$. 
The GSEP field sample includes 132 CCs, 686 RR Lyrae stars,
2819 LPVs, 1377 eclipsing variables, 2 supernovae,  and 9 supernova
candidates in the background sky.  Tile LMC 8\_8 is fully covered by the OGLE~IV GSEP, 
of which  it represents the central 1.2 square degrees (hence a portion corresponding 
to about 1/5 of the total GSEP area) and,  according to OGLE IV, it  contains  19 
CCs, all having a VMC counterpart,  and 223 RR Lyrae stars,  of which 
219 have a VMC counterpart (see columns 6 and 7 of Table~\ref{tab:opt_data}).  
EROS-2 covers 
  only 2/3 of tile LMC 8\_8. 
   In the portion where OGLE IV and EROS-2 data 
overlap  it is possible to compare directly the  results obtained by the two 
microlensing surveys. 
Table~\ref{tab:infoCC_GSEP} and 
 Table~\ref{tab:infoRR_OGLEIV_GSEP} 
provide the cross-identification (namely, OGLE IV and EROS-2 identification numbers) and summarize the properties of the Cepheids and RR Lyrae stars
observed in this  region in common between  EROS-2 and  OGLE IV. We also give in the tables the VMC identification number for all  variables that
have a VMC counterpart.  The bottom part of each table gives instead the cross-identification 
between OGLE IV  and  VMC for Cepheids and RR Lyrae stars in the portion  of tile LMC 8\_8 that is not covered by EROS-2.
In the region in common OGLE~IV  detected 16 
CCs (all having  a VMC counterpart),  of which 8 were detected also by EROS-2,  that also identified in this area a further 
source (lmc0381l13722, VMC-J060325.19-663124.4)  that we found to be a double-mode Cepheid with no counterpart in OGLE~IV. 
In the common region there are also two sources with very small amplitude that OGLE~IV classified as ``spotted stars" 
while R12b classified  as CCs according to their position in the infrared PL. 
Furthermore, in the common area OGLE~IV identified 
153 RR Lyrae stars, of 
which 151 have a counterpart in VMC, to compare with the 109 confirmed RR Lyrae stars detected in the same region by EROS-2. 
 We also note that 4 of the 109 RR Lyrae stars for which we confirmed the 
EROS-2 classification do not have a counterpart  in the OGLE~IV  catalogue of variable 
stars in the whole GSEP field (see 
  Table~\ref{tab:infoRR_OGLEIV_GSEP}).
Fig.~\ref{fig:checkP_EROS-2_OGLEIV} 
shows the comparison between P$_{EROS-2}$ and P$_{OGLEIV}$ for 345 
RR Lyrae (black points) stars and 94 CCs (black asterisk)  in 
common between the two surveys in the GSEP area. 
There is general good agreement between the two period determinations, with only a few exceptions: 
LMC570.15.7157,  LMC562.05.10232, LMC562.21.10078 (OGLE~IV Ids);
the F-mode CCs: LMC562.13.11358, LMC570.05.38, LMC562.05.9009, LMC562.04.66, 
LMC563.15.8125; and the FO mode CC: LMC562.27.37. For all of them P$_{OGLEIV}$ is the correct value. 
For the F/FO mode CC LMC562.25.1125,  and for the RRc star LMC562.02.8742
P$_{EROS-2}$ and 0.5 $\times$ P$_{EROS-2}$, respectively, better fit the 
EROS-2 data. 
Of these sources, only  LMC570.15.7157 falls inside 
 the VMC tile LMC 8\_8.

\begin{figure*}
\begin{centering}
\includegraphics[width=8cm, height=8cm]{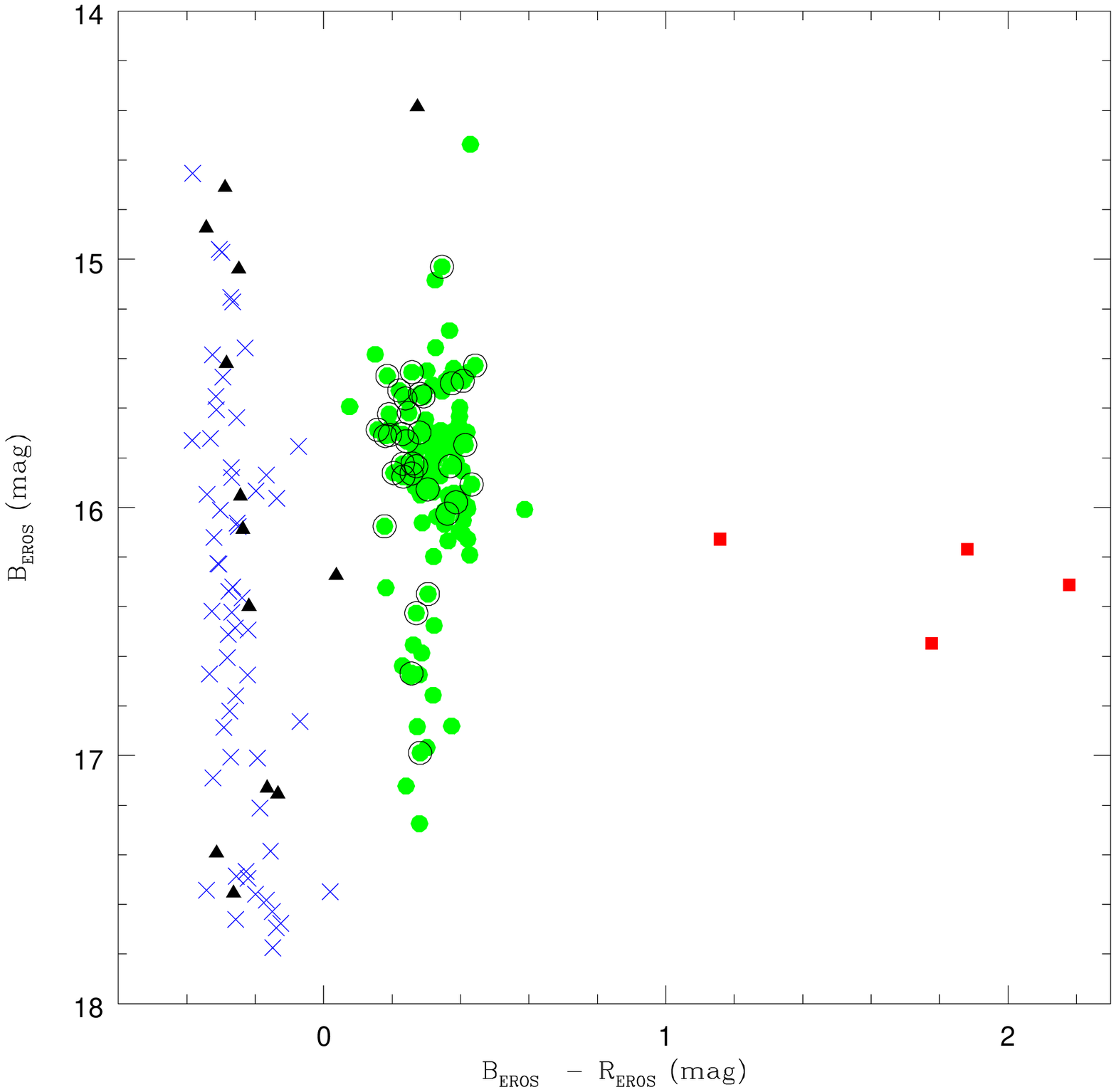}
\includegraphics[width=8cm, height=8cm]{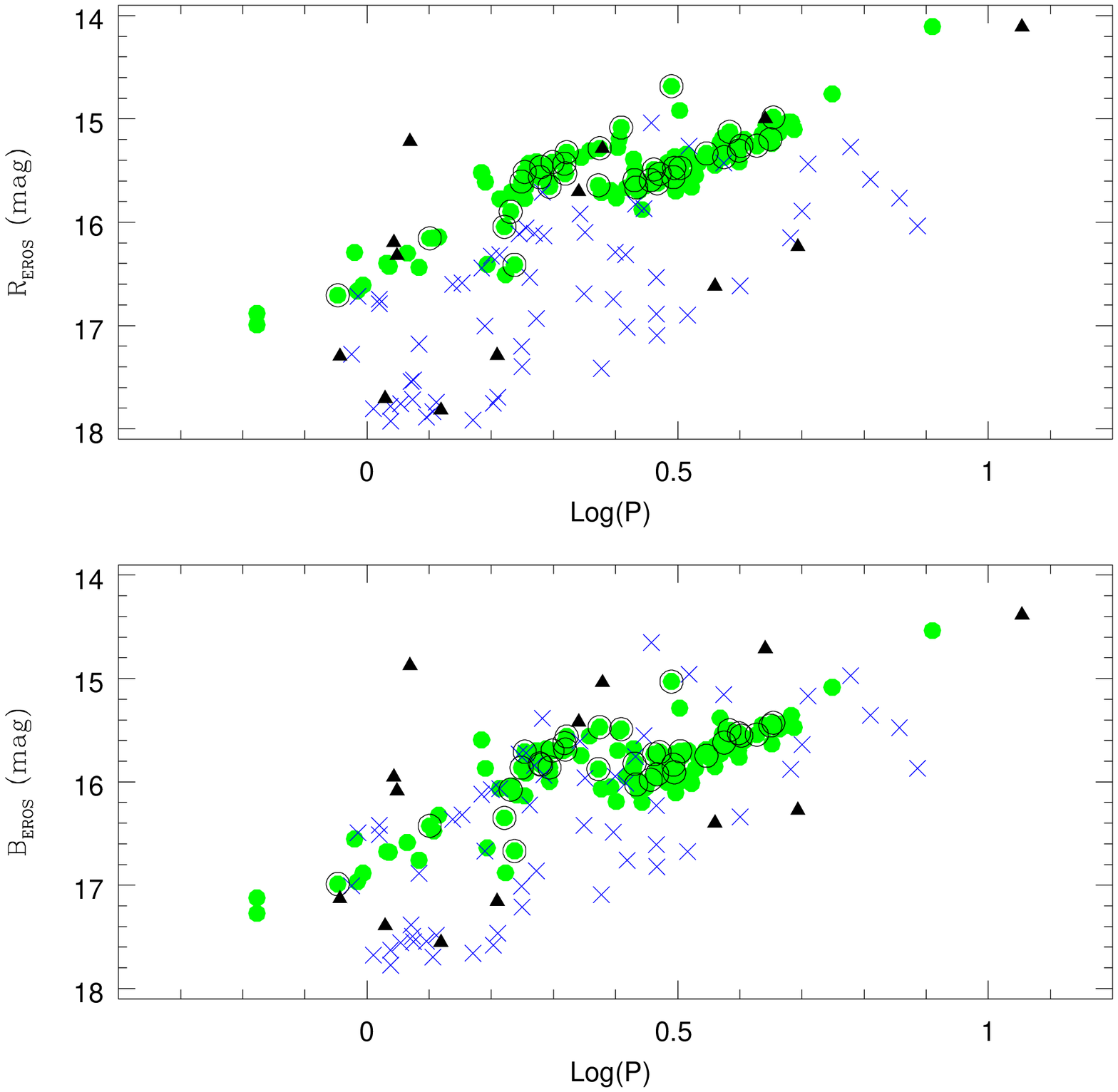}
\caption{Left Panel: $B_{EROS}$, $B_{EROS} - R_{EROS}$ CMD of EROS-2 
CC candidates in tile LMC 8\_3.
Blue crosses, green filled circles and black filled triangles represent 
binary systems, CCs, and small amplitude variables,  respectively. Red filled squares are 
LPVs. Black empty circles are 36 CCs located in the lower portion of tile LMC 8\_3 which is covered also
by OGLE~III, confirming their classification as CCs.
Right Panel: $PL$ relations in the $R_{EROS}$ (upper-right) and $B_{EROS}$ (lower-right)
bands of the EROS-2 CC candidates in tile LMC 8\_3.  
LPVs were omitted. Symbols and colour-coding
are as in the left panel. See text for details.  
}\label{fig:OGLEvsGratis}
\end{centering}
\end{figure*}

\subsection{EROS-2 data in the VMC tile LMC 8\_3}\label{subsec:8_3}
The tile LMC 8\_8 field is not the best place to fine-tune the method
of analysis of  
the CCs, because it contains too few Cepheids. 
Among the LMC outer tiles already completely
observed and catalogued, tile LMC 8\_3 
hosts many CCs (see Table~\ref{tab:opt_data}) and it is much better suited for this purpose. 
Furthermore, 
the lower portion of this tile is covered by the OGLE~III survey (Fig.~\ref{fig:LMCmet}), making possible a direct comparison between  the two surveys.
The EROS-2 catalogues of tile LMC 8\_3  contain respectively 310 
 candidate RR Lyrae stars and 201 candidate  CCs.
We analyzed the optical light curves of all these sources with the GRATIS package. 
For the RR Lyrae stars we also used the LogP, A$_V$  diagram to refine our classification, as described at the end 
of Section~\ref{subsec:gratis},  obtaining in turn a clean 
sample of 262  bona-fide RR Lyrae stars, 
 of which 258  have a VMC counterpart within 1 arcsec.
The visual inspection of the light curves of the  201 candidate Cepheids returned a 
sample of  126 bona-fide CCs of which 125 have  a VMC
counterpart within 1 arcsec\footnote{The candidate CC Id$_{EROS-2}$=lm0310k4094 lies  1.1968657 arcsec from  a possible VMC counterpart, and the comparison of $K_\mathrm{s}$ and EROS-2 optical light curves confirms the
star counteridentification.}. 
 The EROS-2 sample contains  also  58  binaries,  4 candidate LPVs,  and 13 variables  with 
very small amplitudes (generally around 0.1 mag or lower).   As mentioned previously, 
OGLE~III covers only the lower 
1/4 of tile LMC 8\_3 and identified 52 CCs, of which 36 are in common with EROS-2
and 16 do not have a counterpart in the EROS-2 catalogue of CC candidates; 4 of these 16 stars
have a counterpart in the general catalogue of EROS-2 stars but were not classified as CC candidates.
Thus, the total sample of CCs in this tile adds to
142, of which 141 have a counterpart in the VMC catalogue, corresponding to a 99\% completeness of the VMC survey 
 with respect to  the number of Cepheids identified by both EROS-2 and OGLE~III.\\

Tests performed on tiles LMC 8\_3 and 8\_8 show that the CC candidates selected by EROS-2
 using the $PL$ distribution in the right panel of Fig.~\ref{fig:LMCselcc} are mainly  contaminated by binaries. 
 Based on this result we have investigated whether we could  find methods to clean the candidate 
Cepheid sample without checking visually all the light curves, and found that the 
 EROS-2 CMD is well suited for the purpose, as also pointed out by Spano et al. (2011) in their Fig. 8.
The left panel of Fig.~\ref{fig:OGLEvsGratis} shows the 
$B_{EROS}$, $B_{EROS} - R_{EROS}$ CMD of EROS-2 
CC candidates in the region corresponding to tile LMC 8\_3. 
In the CMD  sources classified as binary systems  (ECLs) after visual inspection of the light curve  
are very well separated and definitely  bluer ($B_{EROS} - R_{EROS}  < $ 0.1 mag)  than sources confirmed to be CCs (green points). Furthermore, 
both binaries and CCs appear to be  constrained in small $B_{EROS} - R_{EROS}$
colour intervals in this tile. 
A number of small amplitude variables (black filled triangles in Fig.~\ref{fig:OGLEvsGratis}) also fall  in the region 
occupied by the binaries. As suggested by the amplitudes smaller than
0.1 mag,   the typical periods and the blue colours,
  they likely are a mixture of main sequence (MS) variables like $\beta$ Cepheids, Be stars, slowly-pulsating B variables (see, e.g.  \citealt{balda05} and references therein, for a description of the characteristics of these different types of MS variables). 
Four LPVs (filled red squares) also lie well apart, 
at colours redder than   $B_{EROS} - R_{EROS} \sim$ 1 mag in the CMD of Fig.~\ref{fig:OGLEvsGratis}.
\cite{Spa11} analyzed light curves for 856864 variable stars  in
the EROS-2 database obtaining a final list of 43551 LPVs  in the LMC.
We matched our catalogue of 5800 EROS-2 CC candidates against Spano et al.'s
LPV catalogue and found 296 objects in common. 
 All but one have colours $B_{EROS-2}-R_{EROS-2}
> 1$ mag. 
The 4 LPVs found in tile LMC 8\_3 are all included in the catalogue of LPVs published by \cite{Spa11}. 
The LPVs observed by VMC will be studied further elsewhere.

The right panel of Fig.~\ref{fig:OGLEvsGratis} shows the $PL$ relations in the 
$R_{EROS}$ (upper panel) and $B_{EROS}$ (lower panel) bands of the EROS-2 
CC candidates in tile LMC 8\_3 (the LPVs have been omitted). 
The confirmed CCs are distributed along the two loci occupied by 
first-overtone and fundamental-mode pulsators, respectively. The binaries significantly 
contaminate the Cepheids'  $B_{EROS}$ $PL$, while appear to better separate from Cepheids in the 
$R_{EROS}$ $PL$.  Furthermore,  the bona-fide Cepheids basically remain confined around 
their $PL$s  with only a slightly larger  dispersion  of the  $B_{EROS}$-band relationship, 
the binaries and the vast majority of the small amplitude variables instead shift systematically away 
from the Cepheids going from the $B_{EROS}$  to the $R_{EROS}$ $PL$ and can thus 
easily be disentangled from Cepheids.
In summary, by combining the information from the $B_{EROS}$, $B_{EROS} - R_{EROS}$ CMD and the  
$B_{EROS}$,  $R_{EROS}$ $PL$s it should be possible to separate quite easily
bona-fide CCs from binaries and  small amplitude variables. 

\begin{figure}
\includegraphics[width=8.5cm, height=8.5cm]{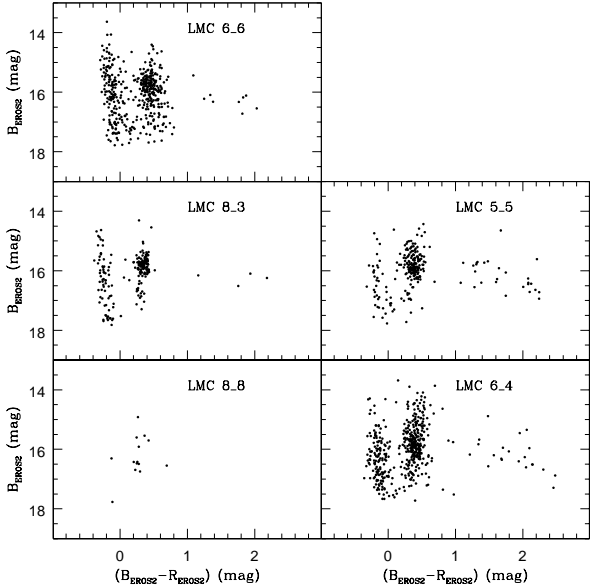}
\caption{
$B_{EROS}$, $B_{EROS} - R_{EROS}$ CMDs of EROS-2 
CC candidates in the 5  LMC tiles completely catalogued as of July 2013. 
Tile numbers are labelled. 
Clearly visible is the separation between binaries, Cepheids and LPVs.}\label{fig:CMDall}
\end{figure}

\subsection{Cleaning the LMC sample of EROS-2 CC candidates}\label{subsec:eros2-cmd}
\begin{figure*}
\begin{centering}
\includegraphics[width=16.5cm, height=16.5cm]{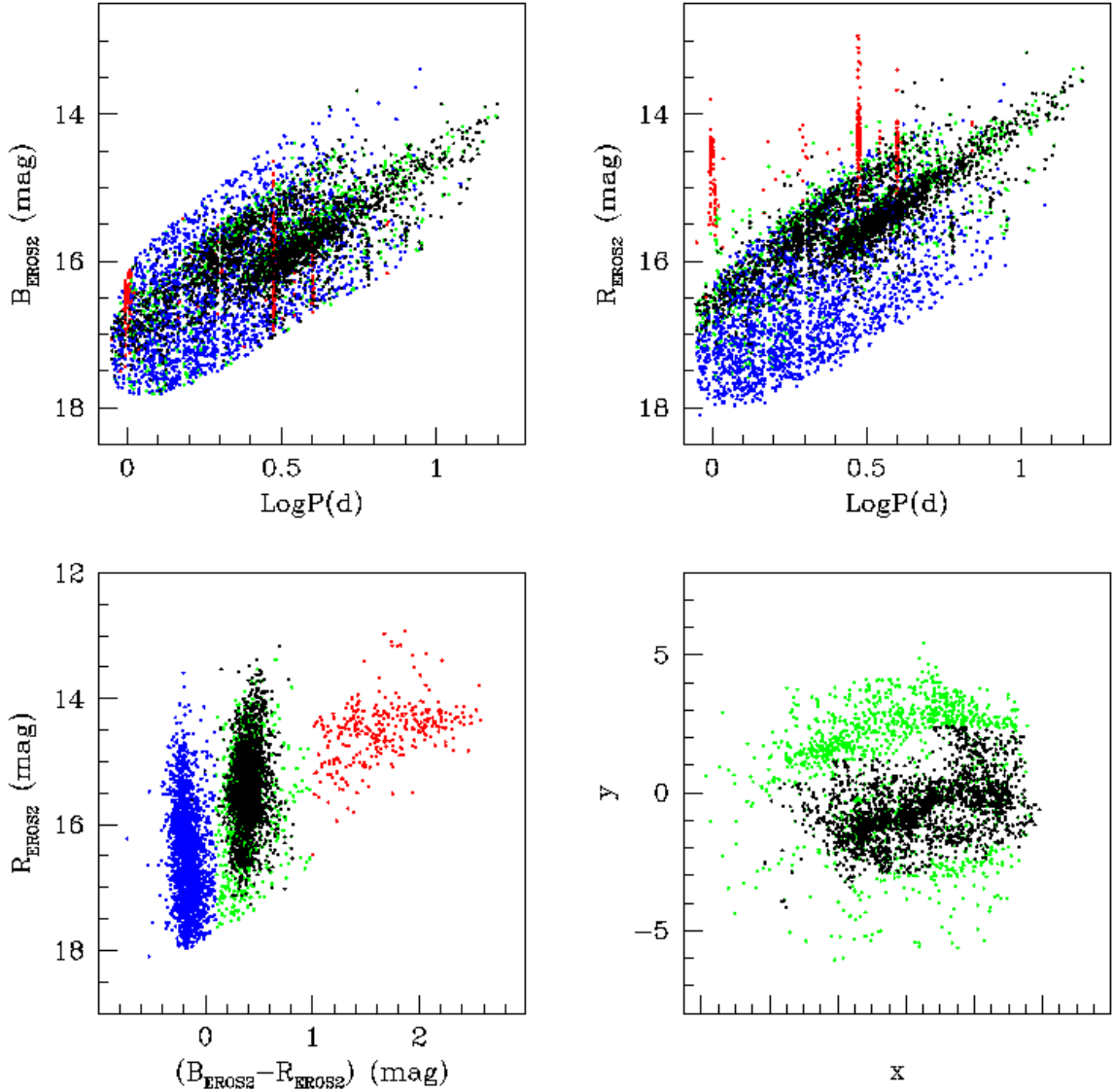}
\caption{
Upper~panels: $PL$ relations in $B_{EROS}$ (left) and $R_{EROS}$ (right) bands 
 of the EROS-2 CC candidates in the LMC. Blue, green and red points mark 
respectively,  ECLs,  CCs and LPVs, according to our
colour-selection criteria. The LPVs fell accidentally in the sample due to the use of aliases of their actual periods 
(correct periods for these stars were published by Spano et al. 2011).
Black points represent CCs in the central part of the LMC
for which a firm classification is  provided by OGLE~III. 
Lower-left~panel: $R_{EROS}$, $B_{EROS} - R_{EROS}$ CMD for EROS-2 candidate 
CCs in the LMC, colour-coding is the same as in the upper panels.
Lower-right~panel: sky distribution of EROS-2 and OGLE~III Classical 
Cepheids in the LMC; colour- coding is the same as in the upper panels.
}\label{fig:cfr4}
\end{centering}
\end{figure*}

Fig.~\ref{fig:CMDall} 
shows the $B_{EROS}$, $B_{EROS} - R_{EROS}$ CMD of the EROS-2 
CC candidates in the 5 VMC LMC tiles 
completely catalogued as of July 2013. 
The separation between binaries, Cepheids and LPVs is clearly visible in all tiles and, as expected, it is more clearcut in the less crowded
tiles. Furthermore, while the distributions of binaries and Cepheids remain sufficiently well separated, their mean colours  are redder in tile
LMC 6\_6 (30 Dor)  due to a large reddening. The range in luminosity 
spanned by the binaries  in the tile LMC 6\_6 also appears to be larger  likely due to the presence  of more massive binaries
in this star forming region. 
As a general rule we expect that  sources with  0.1 $ < (B_{EROS} - R_{EROS}) < 1$ mag are likely to be bona-fide CCs, sources with 
$ (B_{EROS} - R_{EROS} ) < 0.1 $ mag
are likely to be  binary systems, and sources with $ (B_{EROS} - R_{EROS}) \geq 1 $ mag are LPVs. 
According to the afore-mentioned  colour-cuts out of  the 5800 EROS-2 CC candidates  in the LMC 3484 (60.1\%)  very  likely are bona-fide CCs,  2003 (34.5\%) likely are ECLs, and 313 likely are LPVs.  
The results obtained with this procedure 
are shown in Fig.~\ref{fig:cfr4}, where  we have plotted in the lower-left panel the $B_{EROS}$, $B_{EROS} - R_{EROS}$ CMD of the whole sample of EROS-2 CC candidates in the LMC using different colours for the different types of variables,  
and in the upper 
panels the  corresponding $B_{EROS}$, $R_{EROS}$ $PL$s.
Finally, the lower-right panel of Fig.~\ref{fig:cfr4} shows the sky distribution of the LMC CCs selected using the colour-cut in the CMD, they trace very clearly the LMC 
central bar as well as  the overdensity of Cepheids above the bar (see \citealt{Cle11} and discussion in 
Section~\ref{sec:LMC_structure}). 
However, we are aware that the above colour separations may sometimes be too
crude and especially for tiles where the reddening is large and patchy, as for
instance in tile LMC 6\_6, there may be sources with colours between the two 
distributions  that may as well belong  to one or the other groups, and thus 
will need to be checked visually.  

\begin{figure*}
\includegraphics[width=8cm, height=8cm]{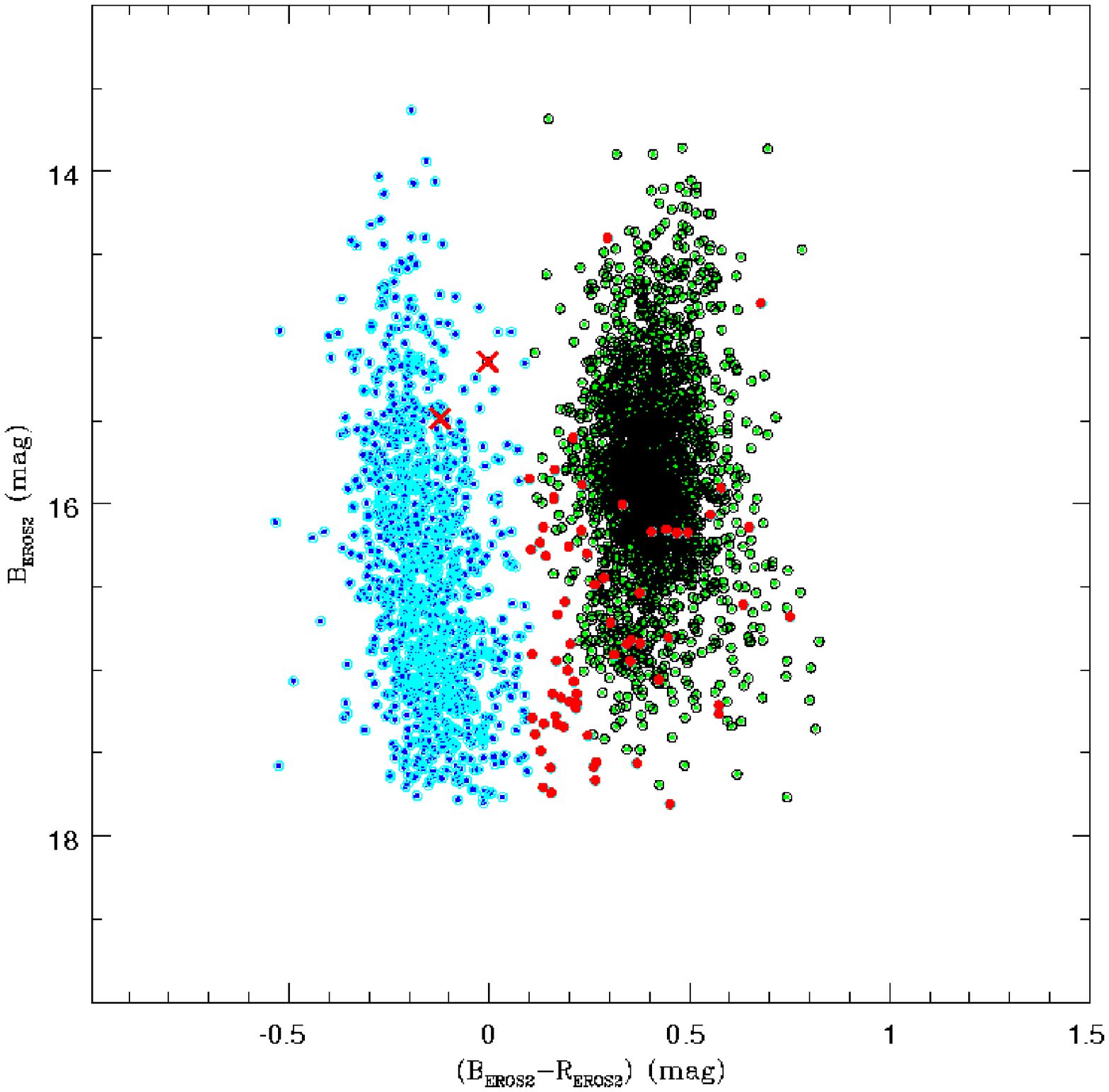}
\includegraphics[width=8cm, height=8cm]{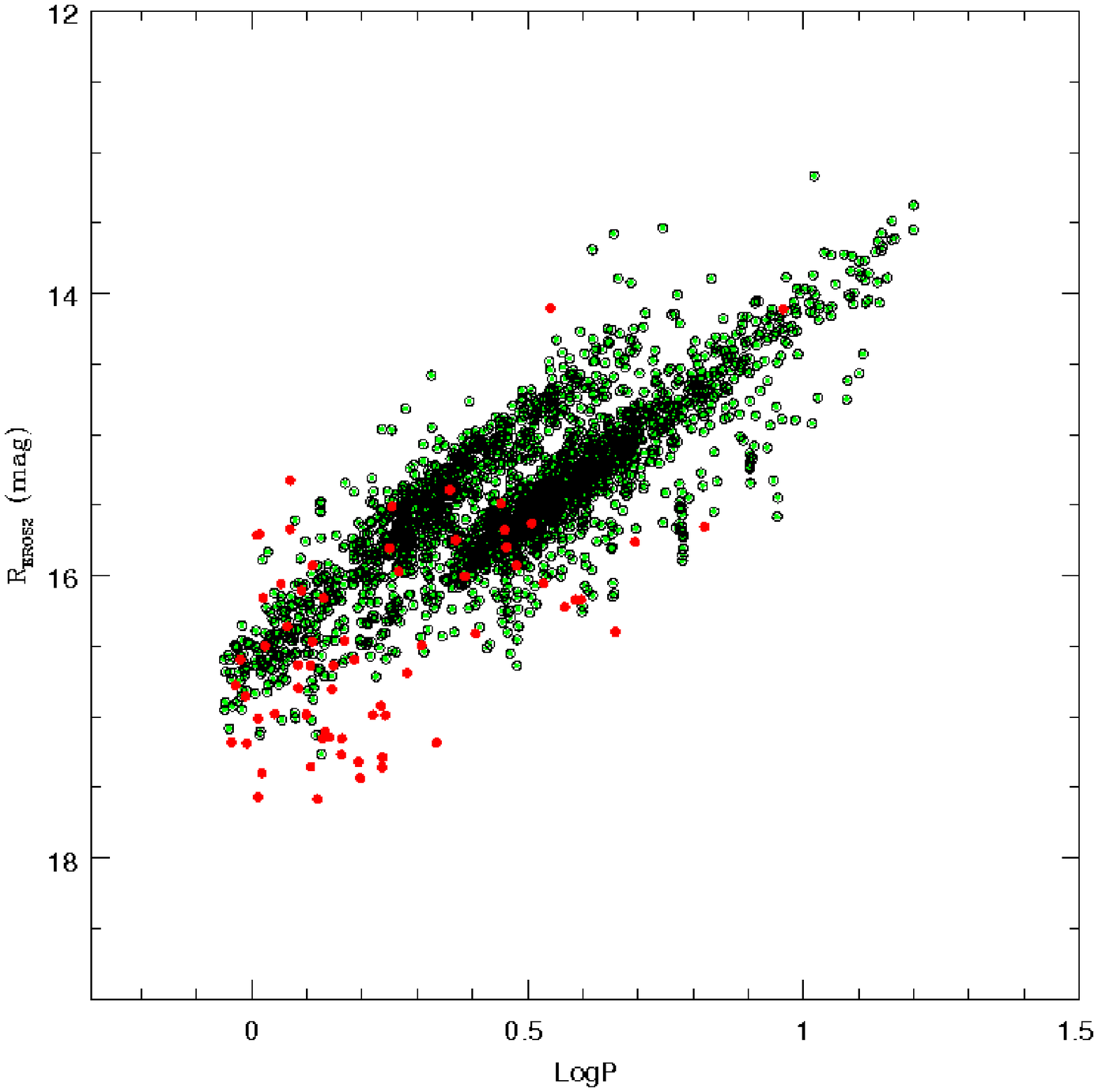}
\caption{Left panel:  CMD of the 3484 EROS-2 CC candidates 
that have a counterpart either in the 
CC or in the  ECL  OGLE~III  catalogues.
Blue points represent EROS-2 CC candidates with colour $B_{EROS-2}-R_{EROS-2}<0.1$mag.
Cyan circles represent EROS-2 CC candidates with an OGLE~III 
counterpart classified as ECL.
Green points represent EROS-2 CC candidates with colour $ 0.1 < (B_{EROS} - R_{EROS} ) < 1.0 $ mag. 
Black circles represent EROS-2 CC candidates with an OGLE~III counterpart 
classified as CC.
 Red crosses mark two CCs, that 
fall in the region of the CMD mainly occupied by binaries. Red filled points 
are binaries falling in the region of the CMD prevalently occupied by Cepheids 
(67 objects). Right panel: $PL$ in the $R_{EROS-2}$ band of the sources with $ 0.1 < (B_{EROS} - R_{EROS} ) < 1.0 $ mag. 
}\label{fig:figura18_cmd}
\end{figure*} 

In order  to  better asses the robustness of our procedures and verify that the  
CCs selected on the basis of the colour-cuts in the CMD are no longer  
contaminated by spurious sources,  we have compared  our selection of the EROS-2 candidate 
CCs in the LMC with the OGLE~III  catalogues of CCs and ECLs  in the central 
region of the LMC.  The EROS-2 catalogue of  LMC CC candidates contains a total number of about 
5800 sources, this number reduces to 5487 if only objects with colour bluer
than 1.0 mag are selected  (i.e. after discarding the LPVs). Of these 5487 objects, 3484 have a counterpart
in the OGLE~III catalogues of CCs and ECLs   when a pairing radius of 1 arcsec is adopted in the cross-match. 
Of the remaining 2003 sources without a counterpart in the OGLE~III catalogues, roughly 1500 lie
outside the FoV covered by OGLE~III. \\
 The left panel of Fig.~\ref{fig:figura18_cmd} shows the 
  CMD (in the EROS-2 bands)  of the 3484 
 objects  with a counterpart in OGLE~III.  This sample contains  2357 CCs and 1062  binaries both according to the colour-cut criteria and the OGLE-III classification. There are only  two objects (red crosses in Fig.~\ref{fig:figura18_cmd}) that we would classify as binaries based on their colour  and    
 are instead CCs according to OGLE~III and the visual inspection of
 the light curves. 
These are stars  lm0551n.20500 and lm0036k.8214 of the EROS catalogue, and 
 correspond to  OGLE-LMC-CEP-0962 and OGLE-LMC-CEP-2595, respectively.
The latter has a very clean light curve, whereas 
 OGLE-LMC-CEP-0962 has variable mean luminosity. As pointed out  in the remarks of the OGLEIII catalogue, 
 the classification  as CC is uncertain.  
 On the other hand, there are 67 sources (red filled points in Fig.~\ref{fig:figura18_cmd})  
that we have classified as Cepheids according to their 
 colour that are instead binaries both  according to 
OGLE~III and the visual inspection of the EROS-2 light curves. 
This corresponds to a 3\% contamination of the Cepheid's sample.
Thirty-three  of these binaries have colour
between 0.1 and 0.2 mag, suggesting that stars with such colours need
visual inspection to be properly classified. 

The right panel of Fig.~\ref{fig:figura18_cmd} 
 shows the $PL$ in the $R_{EROS-2}$ band of the sources with $ 0.1 < (B_{EROS} - R_{EROS} ) < 1.0 $ mag. 
Clearly seen are two separate sequences formed by the fundamental and the first-overtone mode CCs.
Part of the binaries that still contaminate the CC sample (red filled points) 
deviate quite significantly and can be eliminated with a sigma-clipping procedure. 
 This would allow us  to further reduce the residual  3\% contamination of the CC sample.  In conclusion,  the application of colour-cuts in the CMD appears to be a quite robust criterion 
 that  combined with the analysis of the scatter in the  $PL$ relations allows us to extract   a sample of bona-fide CCs more than  97\% clean from contaminating sources.
 A 3\% contamination, in any case is not expected to affect significantly the CC PL$_K$ relations
in regions where only the EROS-2 data are available.

\subsection{Binaries among the  EROS-2 CC candidates}\label{subsec:binaries}
\begin{figure}
\includegraphics[width=8cm, height=8cm]{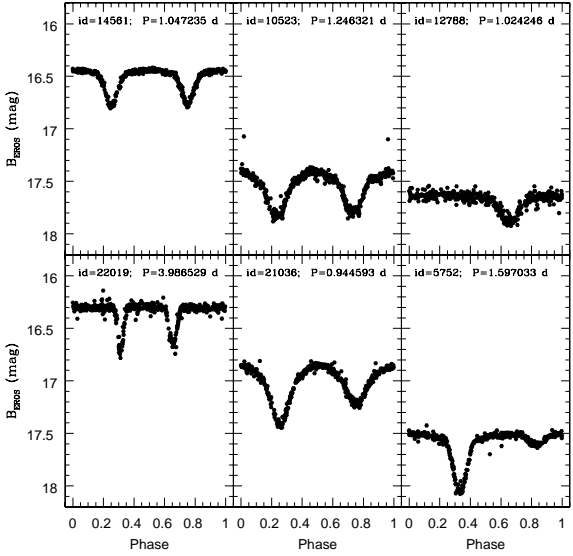}
\caption{Typical light  curves for different types of binaries contaminating the EROS-2 candidate CC sample.}
\label{fig:bin}
\end{figure}

Eclipsing binaries (ECLs) are binary stars where components 
undergo mutual eclipses. The light curves are characterized by periods of
practically constant light with periodic drops in intensity during the
eclipses.
They can be contact, detached and
semidetached binaries containing components of different types 
e.g., giants, dwarfs, MS stars and hence of different age, colours and 
magnitudes; for this reason 
they are not in general used as population tracers. 
The  sample of ECLs  contaminating the EROS-2 catalogue of CC candidates in the 
LMC appears to follow a $PL$ relation similar to the Cepheid 
$PL$ but with a higher dispersion (see Figs.~\ref{fig:LMCselcc}, 
~\ref{fig:OGLEvsGratis} and ~\ref{fig:cfr4}). 
A few examples of light curves, for  different types of 
binaries in our sample, are shown in  Fig.~\ref{fig:bin}. 
A common feature in spite of  those binaries having rather different 
morphological types is their blue colour
(B$_{EROS}$-R$_{EROS} < 0.1 $ mag), hence they appear to be  systems formed by 
pairs of hot main sequence  O, B stars; we then expect that these systems trace a
young population with age $\sim$ 10 Myr (see discussion in 
Section~\ref{sec:LMC_structure}).
Their light curves  often show 
the typical shape of a deformed surface (see, e.g. star 10523 in 
Fig.~\ref{fig:bin}).
To keep a star in contact with the critical lobe, at fixed mass ratio for 
increasing separation  a larger star is needed; since these binaries are on 
a MS this implies a higher luminosity,  
hence there might be a relation between luminosity and period.

In other cases the light curve reveals  a very eccentric detached system, as 
for instance in the case of  star 22019 in Fig.~\ref{fig:bin}, and one could 
make an attempt to compute the
system eccentricity. Other light curves show what probably is a reflection 
effect around the secondary minimum, indicating a large temperature
difference between the components.
We defer a more detailed discussion of our particular sample of hot binaries 
to a further paper  (Muraveva et al., in preparation) where we analyze 
and classify them using the Fourier parameters of the
light curve decomposition.

\section{Comparison with the Star Formation History results}\label{sec:SFH}
\begin{figure}
\includegraphics[width=8cm, height=9cm]{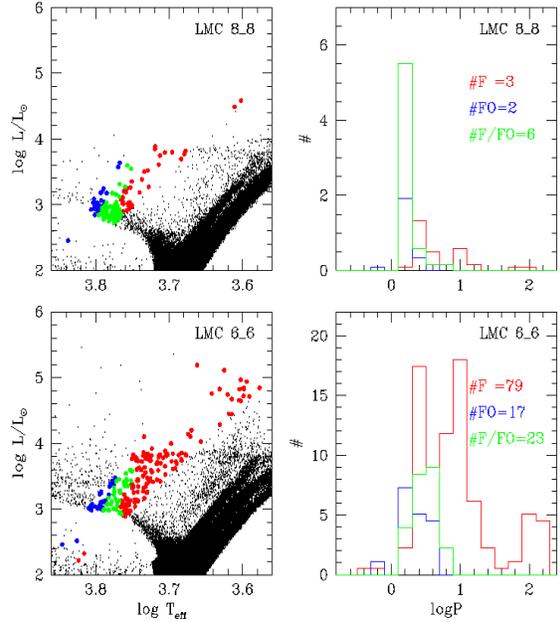}
\caption{Results obtained from the SFH analysis in tiles LMC 8\_8 (upper panels) and 6\_6 (lower panels). Left panels
represent the syntethic CMD with red, blue and green points indicating F, FO and F/FO Classical Cepheids.
Right panels show the distribution of predicted Classical Cepheids with the period; colour coding as in left panels.
}\label{fig:SFH}
\end{figure}

A comparison can  be performed, on a tile-by-tile basis, of the properties and number density of the variable stars with the SFH recovered from the non-variable stars in the same tile.
For instance, the SEP and 30 Dor regions differ significantly not only in terms of number
and type of variables they contain but also in the properties of the parent non-variable stellar populations.
According to 
 the present analysis there are 19 CCs 
(14 FO, 2  F, 2 F/FO, 1 FO/2O, and 1 2O) in the SEP tile, 
and 321 CCs in the 30 Dor region (of which 162 pulsate
in the F mode, 136 pulsate in the FO, 8 are F/FO and 15 are
FO/2O). 
Preliminary simulations indicate that the very different number  of 
Cepheids in these two LMC fields can be explained by the significantly different SFH histories
occurred in the SEP and 30 Dor regions. 
Using VMC data, \cite{Rub12} derive the SFH for 11/12 of the tile LMC 8\_8, 
and for 2 small subregions of the tile LMC 6\_6. They find that the tile LMC 8\_8 has 
only a very tiny fraction of its star formation at ages younger than 200 Myr. 
This lack of young star formation explains the lack of F-mode Cepheids with 
periods longer than logP=0.6 in the SEP field, as well as their concentration 
at even smaller periods (see Fig.~\ref{fig:SFH} upper panels). 
Filtering the best fitting models from Rubele et al. 
(2012) with the theoretical Cepheid instability strips from \cite{Mar04},  
 we expect 9 F and 9 FO Cepheids in this tile; the numbers observed are certainly weighted more 
 toward FO than predicted, but the total number is almost the same.
 The 30Dor field instead is found to have a 
significant star formation activity at all recent ages, which explains the 
large numbers of Cepheids with large periods. If we extrapolate the SFH derived 
from the 2 small subregions to the entire 30Dor tile, we find expected numbers 
of 103 and 39 for F and FO pulsators, respectively (see Fig.~\ref{fig:SFH} lower 
panels). These numbers are smaller 
than the observed ones, especially for the FO Cepheids. We find a 
qualitative agreement between the observed period distributions, and those 
expected from the SFH analysis.  Fewer Cepheids are predicted by the models, 
however, this could be due to the partial inactivity, for stars younger than
0.6 Gyr, 
of the regions studied in \cite{Rub12}. These regions were selected
for their low differential extinction but the
average star formation rate for younger stars in 30 Dor is probably
higher than those measured in the selected areas.

\section{Spatial distribution of the Magellanic variable stars}\label{sec:LMC_structure}
\subsubsection*{EROS-2 data}
\begin{figure*}
\begin{centering}
\includegraphics[width=16.5cm, height=16.5cm]{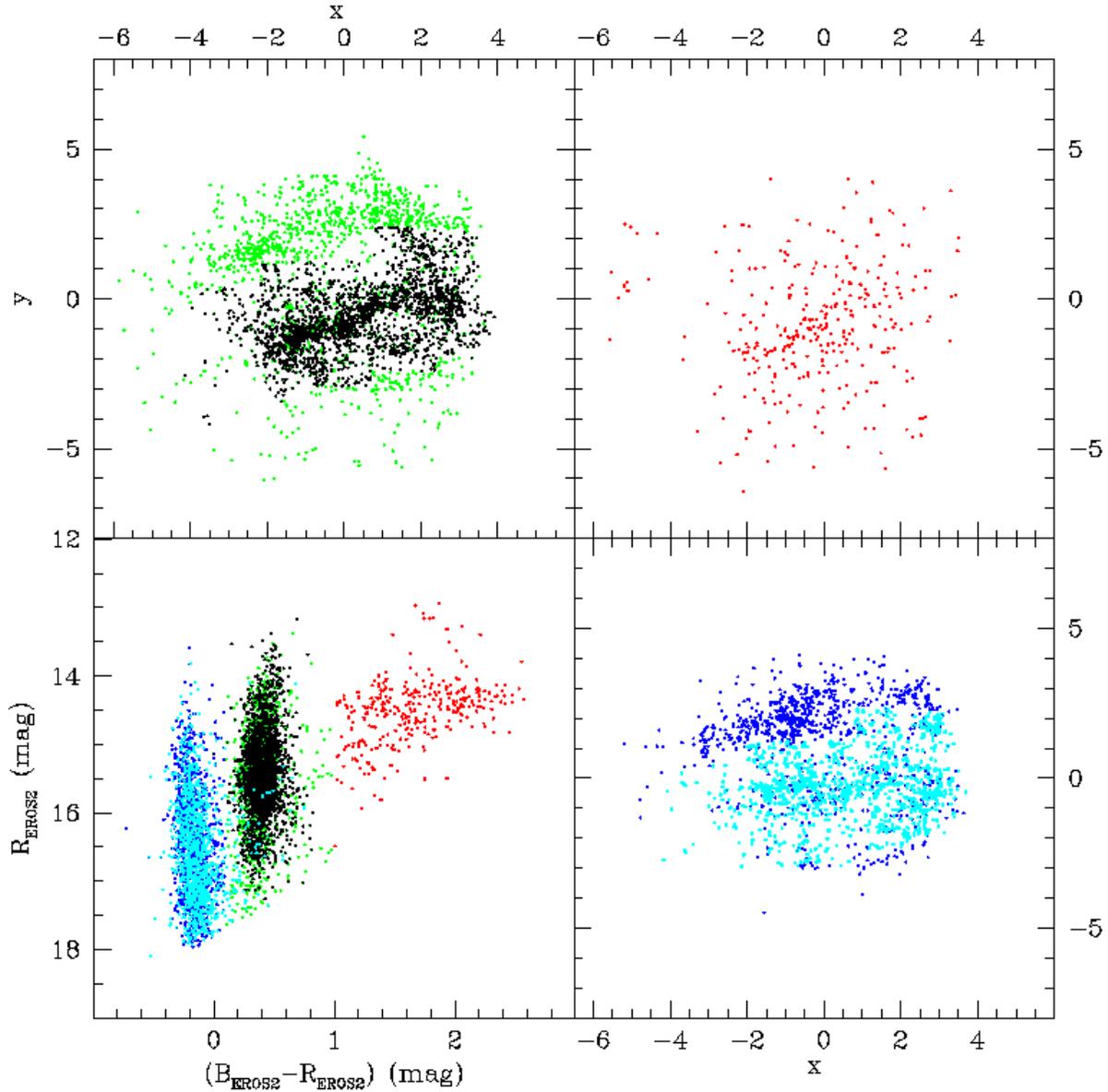}
\caption{Lower-left panel: (R$_{EROS}$, B$_{EROS}$-R$_{EROS}$) CMD of the
 EROS-2 
CC candidates in the LMC. Blue, green and red circles indicate 
respectively hot ECLs, CCs and LPVs,  according to our 
colour selection criteria (see text for details).
Upper-left panel: spatial distribution of the CCs; upper-right panel: distribution of the LPVs 
and binary systems (lower-right panel) that contaminate the list of EROS-2 CC candidates in the LMC.
x and y are related to $\alpha$ and $\delta$ according to equations described in the Appendix.
}
\label{fig:CMDandLMCmapsdue}
\end{centering}
\end{figure*}

Fig.~\ref{fig:CMDandLMCmapsdue} shows the spatial distribution of the three
main  types of variables (CCs, binaries, and LPVs) contained in our  EROS-2 catalogue of candidate
CCs in the LMC.  They trace the 
  spatial distribution of their parent populations,
namely, a young population with the Cepheids (green points) and the hot binaries (blue points), 
and intermediate-age/old stars with the LPVs (red points).
The LPVs, that accidentally fell in our sample (see Section~\ref{subsec:eros2-cmd}), 
spread all over the EROS-2 FoV with a feeble overdensity in the
internal regions; the whole catalogue of the EROS-2 LPV in the LMC is studied by \cite{Spa11}.
Like the CCs,  the hot ECLs are mainly concentrated towards the LMC bar, but show  
a  more structured distribution than the CCs.
These objects are indeed hot young MS stars, which are more
clustered than CCs, and are located in the regions not occupied by CCs.
This could suggest, that the distribution
of hot binaries shows the places of recent ($\sim$10 Myr) star
formation activity. Further investigation about this issue will be
developed in a forthcoming paper (Muraveva et al. in preparation).

\subsubsection*{EROS-2, OGLE~III and OGLE~IV data}
\begin{figure*}
\begin{centering}
\includegraphics[width=8.5cm, height=8.5cm]{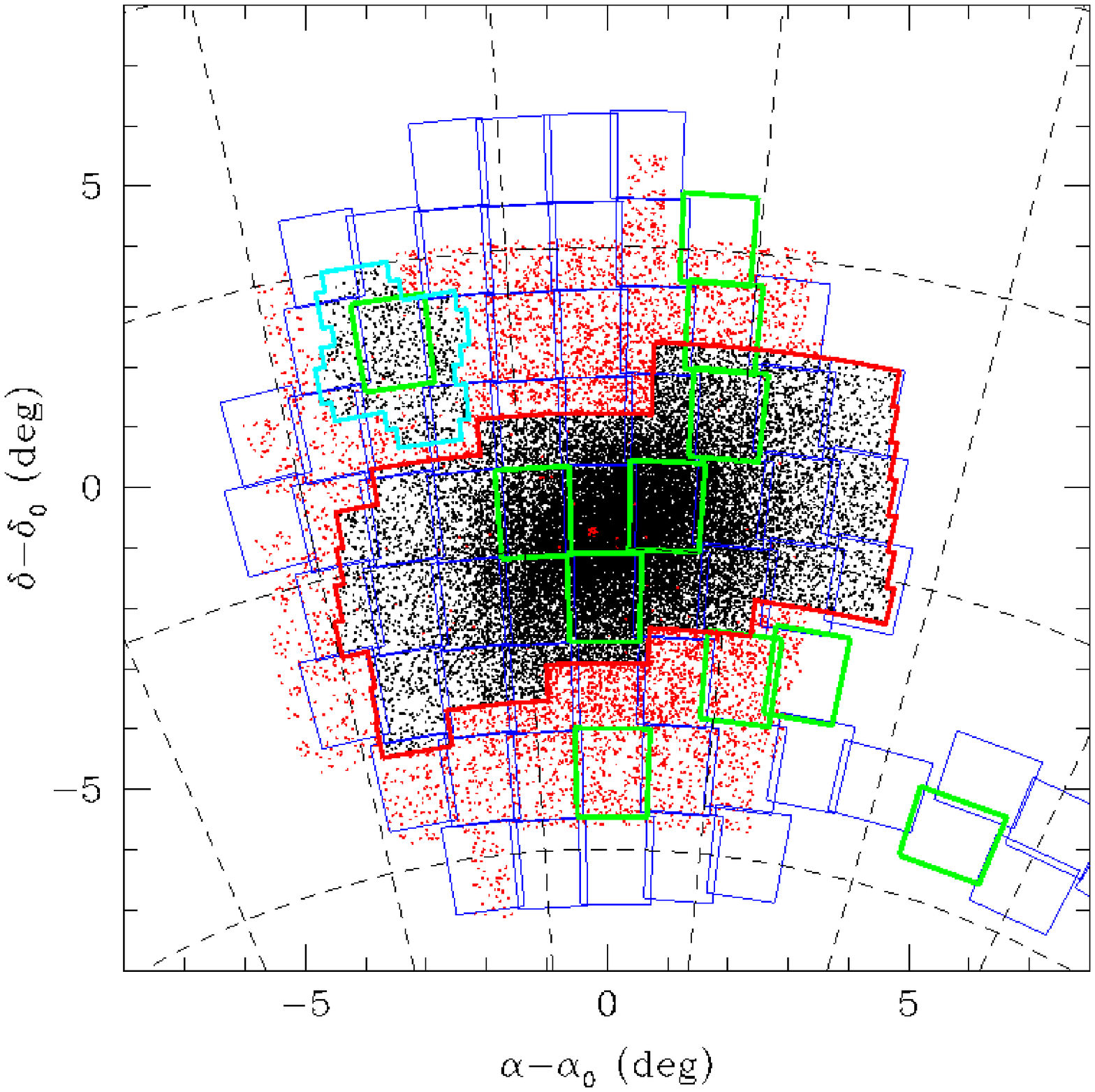}
\includegraphics[width=8.5cm, height=8.5cm]{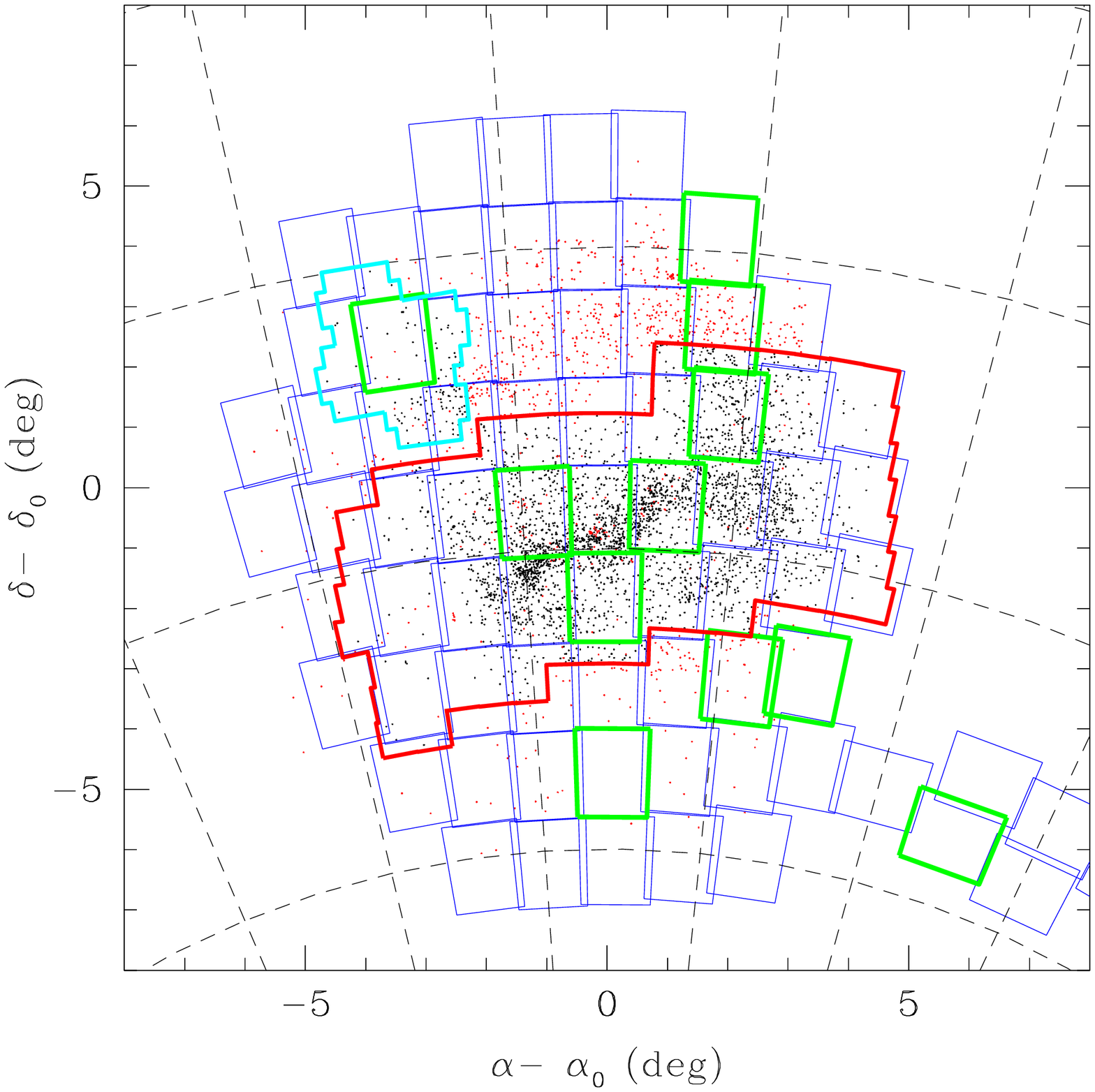}
\includegraphics[width=8.5cm, height=8.5cm]{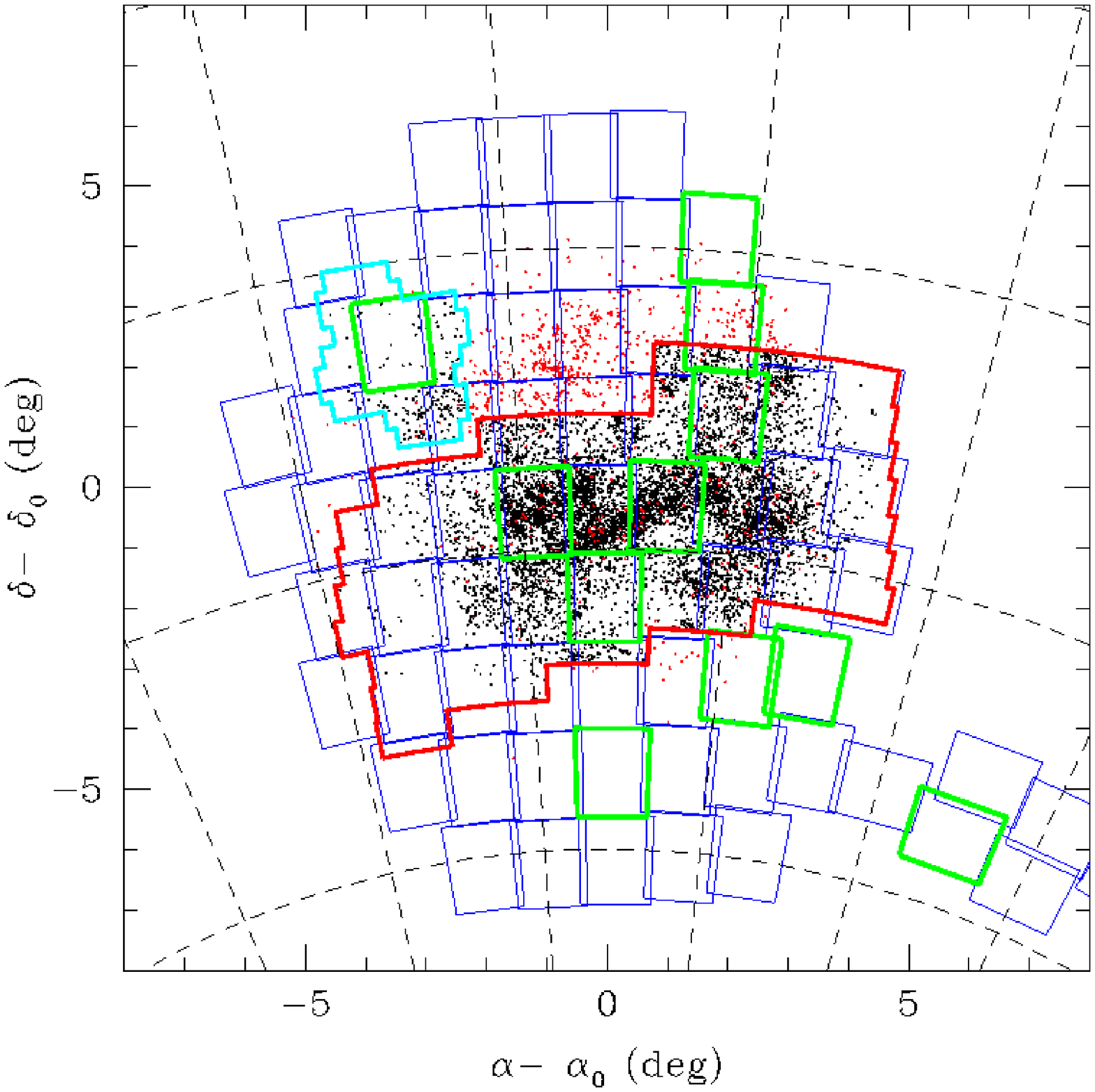}
\caption{Distribution of bona-fide RR Lyrae stars (upper-left), CCs (upper-right), and hot ECLs (lower-left)  in the LMC, according to the OGLE~III, OGLE~IV
and EROS-2 surveys. Black points represent sources with OGLE data; red points
represent EROS-2 confirmed variables of the three different types that do not have an OGLE
counterpart within 5 arcsec. $\alpha_0$=81.0 deg and $\delta_0$=-69.0
deg. Contours are colour coded as in Fig.~\ref{fig:ogleObs}. Green rectangles underlines tiles already observed as of July
2013.
}\label{fig:RR_LMCmap}
\end{centering}
\end{figure*}

Finally, in Fig. ~\ref{fig:RR_LMCmap} we compare  
the structure of the LMC as traced by RR Lyrae stars (upper-left panel), CCs (upper-right panel) and hot 
ECLs (lower panel) obtained by combining results for the LMC variables from the OGLE  
(black points; \citealt{Sos08a,Sos09,Sos12}), and the EROS-2 surveys (red points; 
see Section~\ref{sec:EROS_LMC}). 
To extract the sub-set of hot binaries from OGLE~III general catalogue  of ECLs (Graczyk et al. 2011)  
 we first cross-matched 
the EROS-2 hot binaries  against OGLE ECL catalogue.  In this way we could 
define the region occupied by the hot ECLs in the ($V,V-I$) plane, this corresponds to sources 
with  $V$ magnitudes in the range of 12.5 to 18.5 mag and $V-I$ colours  in the range of  $-$0.35 to  0.4 mag.

These distributions appear to be significantly different. This is not surprising given the 
different stellar populations traced  by these  types of variables (see Introduction). 
The RR Lyrae stars have a larger density  in the central region of the LMC, however,  they are still 
present in the peripheral areas covered almost exclusively by EROS-2. 
Their distribution is smooth and likely traces the
LMC halo. On the contrary, both the CCs and ECLs are strongly concentrated 
towards the LMC bar and seem to almost disappear moving outside the region covered by the
OGLE~III observations (red contour). 
 Fig. ~\ref{fig:RR_LMCmap} shows that the two distributions appear remarkably similar but that, 
as one would expect from Galactic examples, the hot young massive binaries are more sharply concentrated toward
the most recent areas of star formation;
 in particular the 
overdensity of hot binaries near the 30
Dor region, does not have a counterpart in the CC distribution. 
EROS-2, having a much larger  coverage of the LMC than OGLE~III, 
 further highlights  the difference between RR Lyrae star
 and Cepheid distributions  
and also reveals a  feature not captured   
by  OGLE~III because of the smaller FoV: 
as already noted in \cite{Cle11} based on   a preliminary analysis of the EROS-2 data, 
the distribution of the EROS-2 Cepheids clearly confirms 
the existence  in the LMC of an overdensity of CCs displaced about 
2 degrees above the central bar,  
and running  almost parallel to it, to which it  connects at  its western edge.  
Evidences for an  overdensity of CCs  above the 
main bar of the LMC were  already known from the studies of \cite{schmidt-kaler77} 
which  were based on observations
of the LMC CCs by \cite{payne-gaposchkin74} (see Fig. 11 
in \citealt{schmidt-kaler77}) and, more recently, by 
 \cite{nikolaev04}  whose study is based on the MACHO sample of LMC 
CCs (see Fig.~1 of  \citealt{nikolaev04}). They are 
interpreted as the signature of a  northwest spiral arm of the LMC.
The distribution of 
CCs  in the OGLE~IV GSEP field (cyan contours in Fig.~\ref{fig:RR_LMCmap}) also shows an increase in the south-eastern part of the 
field, confirming the overdensity of Cepheids highlighted by the EROS-2 data (\citealt{Cle11}).
A similar overdensity is also seen in the distribution of extended objects (star clusters and associations) in the Magellanic Clouds (\citealt{bica08} and references therein). 
We expect that the VMC data, combined with the microlensing data, both EROS-2 
and the forthcoming OGLE~IV, will
allow us to  revisit and  fully characterize this feature  by better measuring 
its location and displacement along the line of sight by means of our $PL$ 
 relations based on multi-epoch $K_\mathrm{s}$ data.  
This will provide stronger constraints on the theoretical 
models of the LMC structure as well as of the MS formation  since the LMC 
central bar and the Cepheids' overdensity above may be dynamical features that do not resemble 
classical bars/spiral arms  found in spiral galaxies. 

\begin{figure*}
\begin{centering}
\includegraphics[width=8.5cm, height=8.5cm]{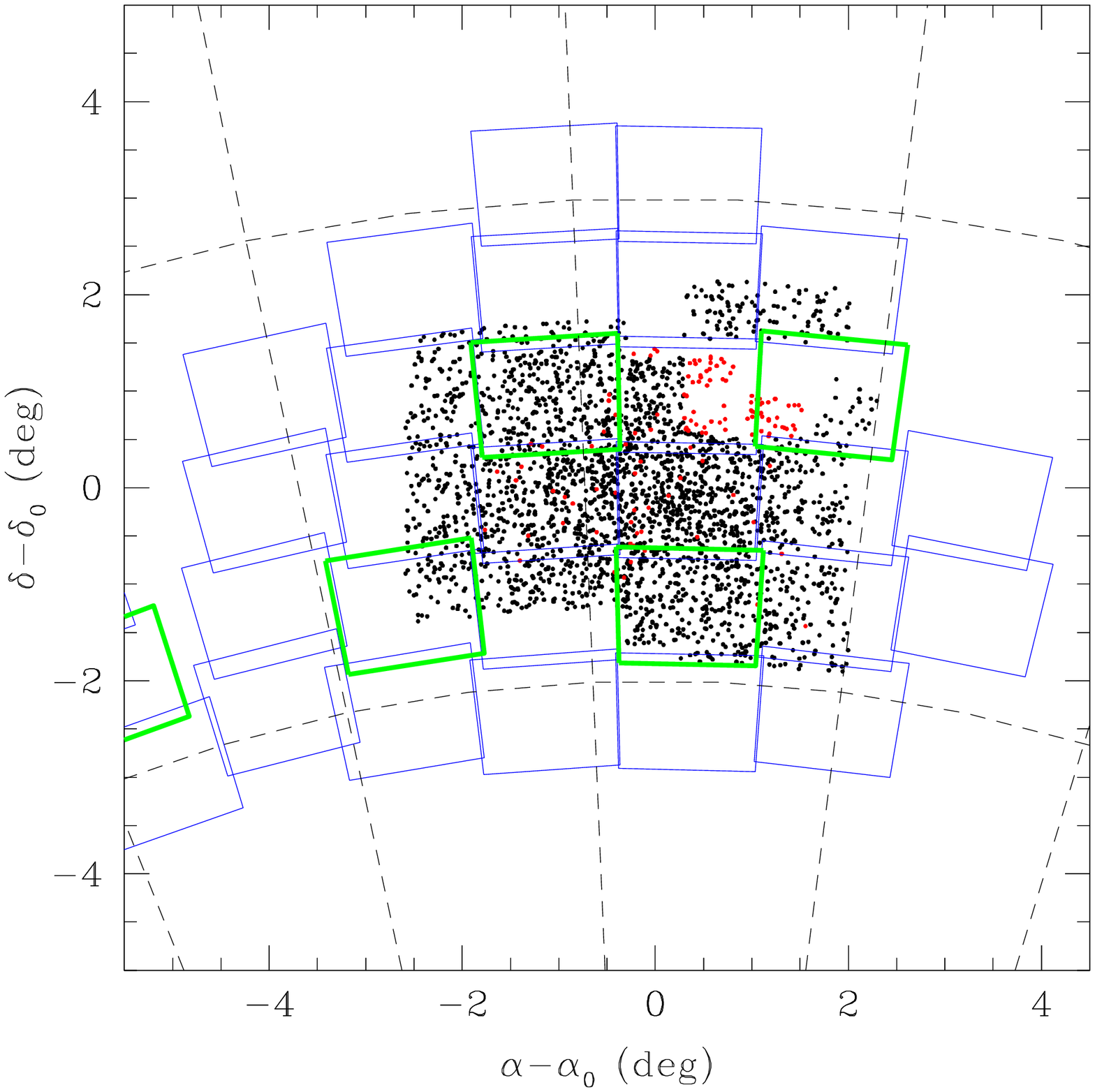}
\includegraphics[width=8.5cm, height=8.5cm]{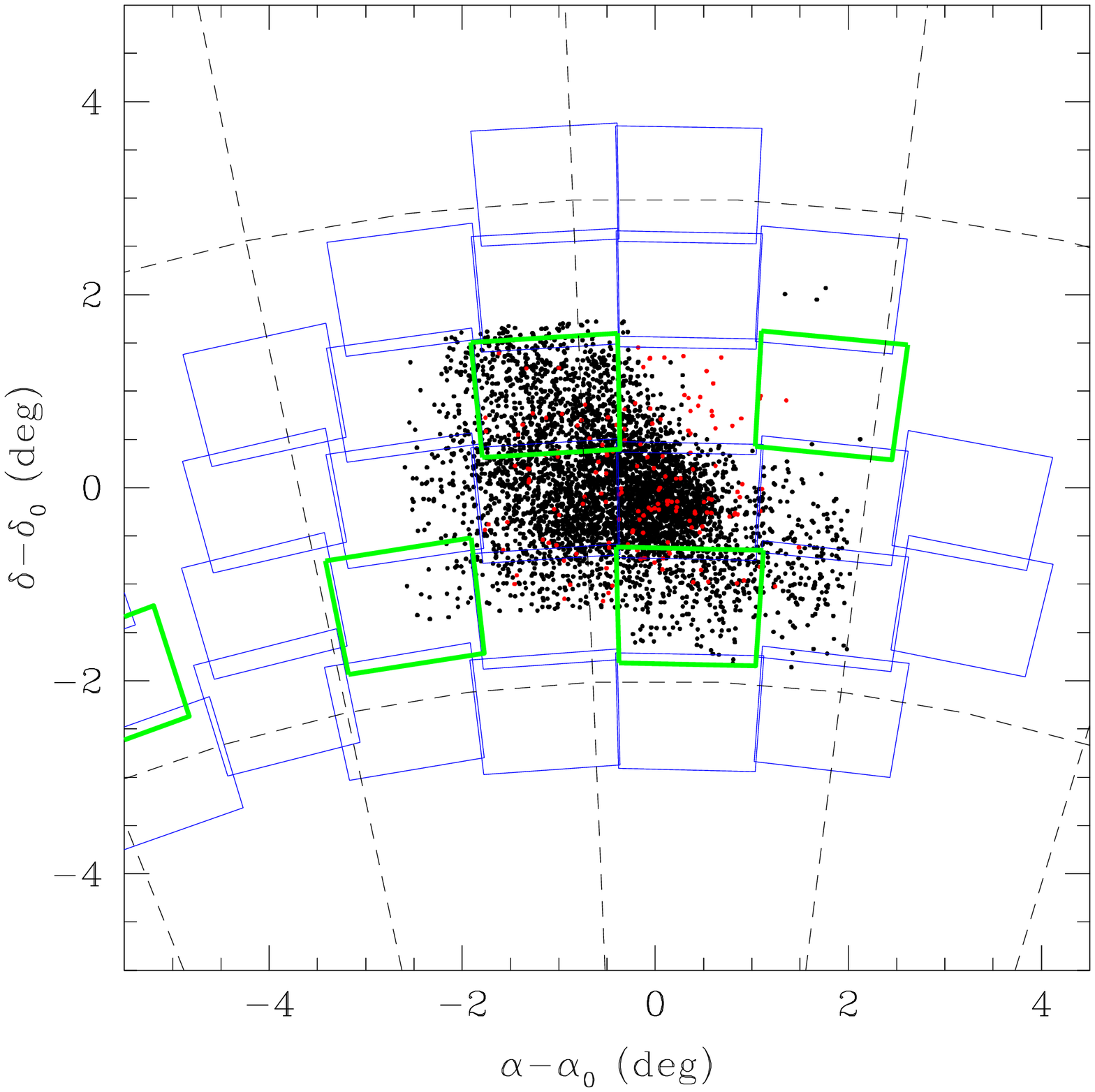}
\caption{Same as Fig.~\ref{fig:RR_LMCmap} but for the 
RR Lyrae stars (left panel) and CC (right panel) in the SMC where 
$\alpha_0 = 12.5$ deg and $\delta_0=-73.0$ deg. 
}
\label{fig:ogleObsSMC}
\end{centering}
\end{figure*}

For completeness, Fig.~\ref{fig:ogleObsSMC} shows the distribution of OGLE (black points) 
and EROS-2 (red points) variables in the SMC area.
As in the LMC  the CCs (right panel) are  
 more clustered in the central 
part of the galaxy, while the RR Lyrae stars (left panel) are 
 more homogeneously distributed, although projection effects make evidence 
less clearcut than in the LMC. On the other hand, in the SMC the number of CCs 
(4630 according to OGLE~III) is about twice the 
number of RR Lyrae stars (2475). This is consistent with the larger gas 
content and more recent star formation in the SMC compared to the LMC.

At present no optical data are available for the area corresponding to the Bridge between the two Clouds. 
However, both STEP@VST and  OGLE~IV have already observed the Bridge 
area and
discovered there both RR Lyrae stars and Cepheids. 
Analysis of the light curves and a full characterization  of the variable stars in these tiles 
is in progress.

\section{Summary and Conclusions}\label{sec:summary}
We have presented the strategy we are  applying to the visual and near-infrared data of 
RR Lyrae stars and Cepheids in the Magellanic System 
that  are being observed by the VMC survey.  
In particular we have presented the properties of the VMC data  for RR Lyrae and Classical Cepheids 
in some already observed fields showing examples of light curve and 
describing the data aquisition. We compared 
the optical (EROS-2, OGLE~III and OGLE~IV) and infrared (VMC) data in 5 LMC 
fields studying the completeness of the VMC survey with respect to the optical
catalogues. We 
analyzed the optical light curves for CC and RR Lyrae stars in the tile
LMC 8\_8 using both the EROS-2 and OGLE~IV data. Comparison between the expected
numbers of CC from the SFH studies and the observed data is performed in two
LMC tiles.
Finally we compared the distribution in RA and DEC of both variable stars.
The analysis described in this paper clearly  shows  that RR Lyrae stars and 
Cepheids sample regions 
(halo, bar, spiral arms) of the MS that are  differently located in space. 
We are now reaching the level of accuracy and detail
to disentangle the fine structure of the Magellanic System and thus separate 
systematics from geometrical effects that 
influence these primary indicators of the cosmic distance  ladder.

\section*{Acknowledgments}
M. I. M. thanks the Royal Astronomical Society and the University of
Bologna 
for grants to  spend five months at the University of Hertfordshire during the spring-summer 2010. 
C. Maceroni and M. Cignoni are warmly thanked for their help with the interpretation of the 
binaries contaminating the Cepheid sample.
M. I. M. wishes to thank S. Leccia, R. Molinaro and F. Cusano,
 for useful scientific and technical discussions.
This work made use of EROS-2 data, which were kindly provided by the EROS 
collaboration. The EROS (Exp\'erience pour la Recherche d'Objets Sombres) 
project was funded by the CEA and the IN2P3 and INSU CNRS institutes.
We acknowledge the OGLE team for making public their catalogues.
We thank the UKs VISTA Data Flow System comprising the VISTA pipeline at the Cambridge Astronomy Survey 
Unit (CASU) and the VISTA Science Archive at Wide Field Astronomy 
Unit (Ed- inburgh) (WFAU) for providing calibrated data products under the support of the STFC.
Financial support for this work was provided by PRIN-INAF 2008 (P.I. M. Marconi) and by COFIS ASI- INAF I/016/07/0 (P.I. M. Tosi).
RdG acknowledges partial research support from the National Natural Science Foundation of China (NSFC)
through grant 11073001.
R.G. is a Postdoctoral Fellow - Pegasus of the Fonds Wetenschappelijk Onderzoek (FWO) - Flanders.

\label{lastpage}

\linespread{0.5}
\begin{table*}
\tiny 
\caption{
Counteridentification (VMC - OGLE~IV - EROS-2) and properties of the Cepheids  in the LMC 8\_8 tile (SEP; see text for details).}\label{tab:infoCC_GSEP}
\begin{tabular}{llcclllcc}
\hline
~~~~~~~~~~~~VMC\_ID	 &    OGLE~IV\_ID	       &  Type               & Mode               &   P$_{\rm OGLE~IV}$      &  EROS\_ID	     & P$_{\rm EROS}$  &  Type     &    Mode   \\ 
                                                &                                        &  OGLE~IV      &  OGLE~IV        &     ~~~~(day)                 &                                   &   ~~(day)           &  R12b+t.w. & R12b+t.w. \\
\hline
VMC-J055406.28-661228.2  &    LMC563.21.27	 &  DCEP  & F/FO    &     ~~3.41944    &   ~~~~~~ $-$	     &  ~~~~$-$ & $-$ & $-$\\ 
VMC-J055432.96-664223.0  &    LMC563.04.13	 &  DCEP  & FO       &     ~~4.9929       &   ~~~~~~ $-$     &  ~~~~$-$ & $-$ & $-$\\ 
VMC-J055530.29-660557.7  &    LMC563.20.38	 &  DCEP  & F	       &     ~~3.870411   &   lm0507l6104 &  3.87046 & DCEP& F  \\ 
VMC-J055555.41-663609.8  &    LMC563.03.7925	 &  DCEP  & FO       &     ~~1.481971   &   ~~~~~~ $-$	 &  ~~~~$-$ & $-$ & $-$\\ 
VMC-J055613.39-662234.0  &    LMC563.10.6631	 &  DCEP  & FO       &     ~~1.027533   &   lm0371k18509&  1.027683& DCEP& FO\\ 
VMC-J055635.79-654742.0  &    LMC563.27.87	 &  DCEP  & FO       &     ~~1.188903   &   lm0505n6123 &  1.188733& DCEP& FO\\ 
VMC-J055638.35-660302.7  &    LMC563.19.6609	 &  DCEP  & FO       &     ~~1.214799   &   lm0507m20129&  1.214786& DCEP& FO\\ 
VMC-J055657.28-660732.0  &    LMC563.19.84	 &  DCEP  & FO       &     ~~0.863285   &   ~~~~~~ $-$	 &  ~~~~$-$ & $-$ & $-$\\ 
VMC-J055709.22-655129.4  &    LMC563.27.66	 &  DCEP  & FO       &     ~~0.841186   &   ~~~~~~ $-$	 &  ~~~~$-$ & $-$ & $-$\\ 
VMC-J055711.14-655116.0  &    LMC563.27.67	 &  DCEP  & FO       &     ~~1.044417   &   lm0505n10348&  1.044436& DCEP& FO\\ 
VMC-J055721.61-655125.6  &    LMC563.27.27	 &  DCEP  & 2O       &     ~~1.45545    &   ~~~~~~ $-$	 &  ~~~~$-$ & $-$ & $-$\\ 
VMC-J055922.16-665709.8  &    LMC570.16.6654	 &  DCEP  & FO       &     ~~1.683923   &   lm0382l10126&  1.683674& DCEP& FO\\ 
VMC-J055942.93-670346.8  &    LMC562.26.40	 &  DCEP  & FO       &     ~~1.907626   &   lm0384k15682&  1.907595& DCEP& FO\\ 
VMC-J060318.83-665244.3  &    LMC570.14.6292	 &  DCEP  & FO       &     ~~3.228064   &   lm0383l20185&  3.227865& DCEP& FO\\ 
VMC-J060415.65-663933.8  &    LMC570.22.5193	 &  DCEP  & FO       &     ~~0.87984    &   ~~~~~~ $-$	 &  ~~~~$-$ & $-$ & $-$\\ 
VMC-J060532.15-665638.4  &    LMC570.12.6008	 &  DCEP  & F	     &     ~~3.083327   &   ~~~~~~ $-$	 &  ~~~~$-$ & $-$ & $-$\\ 
VMC-J060325.19-663124.4  &   ~~~~~~~~~ $-$	 &  $-$   & $-$      &     ~~~~~$-$     &   lm0381l13722&  1.277076$^{a}$ & DCEP& F/FO\\
VMC-J055535.45-670217.3  &    LMC562.29.61	 &  OTHER & Spots    &	   ~~3.8280     &   lm0375k12090&  3.902331& DCEP&n.c.$^{b}$\\ 
VMC-J060117.37-665319.5  &    LMC570.15.6150	 &  OTHER & Spots    &	   ~~4.0807     &   lm0382n17908&  4.085779& DCEP&n.c.$^{b}$\\ 
VMC-J055924.79-662930.0  &    LMC563.08.60	 &  ECL   & $-$      &     ~~1.2385056  &  lm0380l11651  &  1.238489& ECL & $-$  \\ 
VMC-J055711.52-664418.6  &    LMC563.02.523	 &  ECL   & $-$      &     ~~1.207677	&   lm0373m20395&  1.207672& ECL &$-$   \\  		     
VMC-J055518.02-670541.2  &    LMC562.29.97	 &  ECL   & $-$      &     ~~2.158649	&   lm0375k19793 &  2.158691& ECL & $-$  \\  
VMC-J060011.48-661037.3 &    LMC571.06.49	 &  DCEP  & FO       &     ~~1.90468    &   ~~~~~~ $-$	 &  ~~~~$-$ & $-$ & $-$\\ 
VMC-J060322.75-660145.9 &    LMC571.04.5309	 &  DCEP  & FO/2O    &     ~~0.708377   &   ~~~~~~ $-$	 &  ~~~~$-$ & $-$ & $-$\\ 
VMC-J060433.05-660915.3 &    LMC571.04.11	 &  DCEP  & FO       &     ~~4.85746    &   ~~~~~~ $-$	 &  ~~~~$-$ & $-$ & $-$\\ 
\hline
\end{tabular}
\begin{flushleft}
$^{a}$ \footnotesize{The analysis of the EROS-2 light curves  revealed that this is a double-mode Cepheid with a secondary period of 
1.768788 days, in good agreement with \cite{Mar09}.}\\
$^{b}$ \footnotesize{The periods and luminosities place these two variable stars in the Cepheid domain and thus they were classified as such in R12. 
However, the EROS-2 light curves show a very small amplitude of the order 
of about 0.1 mag. \\
}
\end{flushleft}
\end{table*}
\linespread{1.3}

\linespread{0.5}
\begin{table*}
\begin{center}
\caption{
Counteridentification (VMC - OGLE~IV -  EROS-2)  and properties of the RR Lyrae stars  in the 
VMC LMC 8\_8 tile (SEP;  see text for details). This Table
is published in its entirety only in the electronic edition of the Journal.}
\label{tab:infoRR_OGLEIV_GSEP}
\begin{tabular}{lllllll}
\hline
~~~~~~~~~~~~VMC\_ID	 &     ~~OGLE~IV\_ID   &  Type          &   P\_OGLE~IV            & ~~~EROS\_ID	 &  ~~Type                        & P\_EROS    \\  
                                                &                                     &OGLE~IV    &~~~~~~(day)            &                                    &      This work                & ~~~(day) \\
\hline
\hline
VMC-J055330.63-662350.3 &   LMC563.12.1169 &  ~~~~~RRab 	     & 0.5561077   & lm0370m20974 &  ~~~~~RRab     &  ~0.556178  \\
VMC-J055342.51-663748.3 &   LMC563.04.8357 &  ~~~~~RRab 	     & 0.729109    & lm0372m4284  &  ~~~~~RRab     &	  ~0.729118  \\
VMC-J055346.14-662741.1 &   LMC563.12.996  &  ~~~~~RRc  	     & 0.3849272   & lm0370n8721  &  ~~~~~RRc      &	  ~0.38492   \\
VMC-J055348.75-664012.1 &   LMC563.04.8711 &  ~~~~~RRab 	     & 0.6528193   & ~~~~~~~$-$ 	  &  ~~~~~~~$-$   &    ~~~~~$-$   \\
VMC-J055352.14-663543.3 &   LMC563.04.8914 &  ~~~~~RRc  	     & 0.3544958   & lm0370n10812 &  ~~~~~RRc    &	~0.354487  \\
VMC-J055410.19-665926.6 &   LMC562.30.1447 &  ~~~~~RRab 	     & 0.617039    & lm0374m5838  &  ~~~~~RRab   &	~0.617036  \\
VMC-J055419.33-662744.6 &   LMC563.12.994  &  ~~~~~RRab 	     & 0.652377    & lm0370n8898  &  ~~~~~RRab   &	~0.652404  \\
VMC-J055421.20-662324.8 &   LMC563.12.1192 &  ~~~~~RRc  	     & 0.3313268   & lm0370m20171 &  ~~~~~RRc    &	~0.331325  \\
VMC-J055431.93-670253.3 &   LMC562.30.1223 &  ~~~~~RRab 	     & 0.5059999   & lm0374m13678 &  ~~~~~RRab   &	~0.505996  \\
VMC-J055433.20-664324.0 &   LMC563.04.1161 &  ~~~~~RRab 	     & 0.640376    & lm0372m13082 &  ~~~~~RRab   &	~0.640384  \\
\hline  				     
\end{tabular}			     
\end{center}
\begin{flushleft}
$^{a}$ \footnotesize{According to OGLE~IV  star LMC563.09.6173 is a Galactic RR Lyrae star.}\\
$^{b}$ \footnotesize{EROS-2 period of the star is 1.328165 days, 
which is an alias of the actual period. 
For this star we list in the P$_{\rm EROS}$ column the period
of 0.6640825 days we derived from our analysis of the star light 
curve with GRATIS.}
\end{flushleft}
\end{table*}			     
\linespread{1.3}

\linespread{1.0}			     
\normalsize
\section*{Appendix}
In this paper we often had to transform the $\alpha$, $\delta$ coordinates into Cartesian
 (x, y) coordinates. This was accomplished using the equations:  
\begin{equation}
{\rm x}(\alpha, \delta)=\rho~ {\rm cos}\phi; ~~~~~~  {\rm y}(\alpha, \delta)=\rho~ {\rm sin}\phi.
\end{equation}
where, when the origin is fixed,  $\rho$, $\phi$ are uniquely 
defined as function of $\alpha$, $\delta$ through the relations:
\begin{equation}
{\rm cos}\rho={\rm cos}\delta ~ {\rm cos}\delta_0 ~ {\rm cos}(\alpha - \alpha_0) + {\rm sin}\delta ~sin\delta_0,
\end{equation}
\begin{equation}
{\rm sin}\rho ~{\rm cos}\phi=-{\rm cos}\delta~ {\rm sin}(\alpha-\alpha_0),
\end{equation}
\begin{equation}
sin\rho~ {\rm sin}\phi={\rm sin}\delta ~{\rm cos}\delta_0-{\rm
  cos}\delta~ {\rm sin}\delta_0 ~ {\rm cos}(\alpha-\alpha_0).
\end{equation}
(see \citealt{van01} and references therein) and, in the case of 
the LMC, $\alpha_0=81$ deg and $\delta_0=-69.0$ deg.\\
For the study of completeness we also took into account the position angle of each tile using the rotation equations:
\begin{equation}
{\rm x1}= {\rm x~cos}[(90+angle) ~\phi/180] ~-~ {\rm y~sin} [(90+angle) ~\phi/180]
\end{equation}
\begin{equation}
{\rm y1} = {\rm x~sin}[(90+angle) ~\phi/180] ~+~{\rm x~cos} [(90+angle) ~\phi/180]
\end{equation}
where the position angles used for tiles LMC 5\_5, 6\_4, 6\_6, 8\_3 and 8\_8 are, respectively, $-$92.6525, $-$95.3605, $-$89.5708,
$-$97.2489, and $-$84.4802 deg,  and the rotation is performed with respect to the centre coordinates of each tile as published in 
Paper I. 
Using this criterium some of the selected
  stars lie in the overlap regions between tiles. 
In the case of SMC maps we used $\alpha_0=12.5$ deg and $\delta_0=-73.0$ deg while 
for the Bridge area we adopted $\alpha_0=44.4$ deg and $\delta_0=-73.2$

\end{document}